\documentclass[]{aa501}

\usepackage{graphics}

\newcommand{\beq}{\begin{equation}}
\newcommand{\eeq}{\end{equation}}
\newcommand{\beqa}{\begin{eqnarray}}
\newcommand{\eeqa}{\end{eqnarray}}

\newcommand{\figpsa}[1]{\resizebox{\hsize}{!}{\rotatebox{0}{\includegraphics{#1}}}}   
\newcommand{\figpsax}[1]{\resizebox{15cm}{!}{\rotatebox{0}{\includegraphics{#1}}}}   
\newcommand{\figpsaa}[1]{\resizebox{\hsize}{!}{\rotatebox{0}{\includegraphics{#1}}}}   
\newcommand{\figpsab}[1]{\resizebox{8.5cm}{!}{\rotatebox{0}{\includegraphics{#1}}}}

\newcommand{\figpswbig}[1]{\resizebox{17.5cm}{!}{\rotatebox{0}{\includegraphics{#1}}}} 
\newcommand{\figpswh}[1]{\resizebox{8.5cm}{!}{\rotatebox{90}{\includegraphics{#1}}}}

\begin{document}

\title{Detailed analysis of Balmer lines in cool dwarf stars
\thanks{Based on observations collected at the Isaac Newton Telescope, La Palma, Spain, and McDonald Observatory, Texas, USA.}} 

\author{P. S. Barklem\inst{1}  \and H. C. Stempels\inst{1} \and C. Allende Prieto\inst{2} \and O. P. Kochukhov\inst{1} \and N. Piskunov\inst{1} \and B. J. O'Mara\inst{3} }
 
\offprints{P. S. Barklem,
}

\institute{Department of Astronomy and Space Physics, Uppsala University, Box 515, S 751-20 Uppsala, Sweden \and McDonald Observatory and Department of Astronomy, University of Texas, Austin, TX 78712-1083, USA \and Department of Physics, The University of Queensland, St Lucia, 4072, Australia}

\date{Received 30/11/01 / Accepted 28/01/02}

\abstract{An analysis of H$\alpha$ and H$\beta$ spectra in a sample of 30 cool dwarf and subgiant stars is presented using MARCS model atmospheres based on the most recent calculations of the line opacities.  A detailed quantitative comparison of the solar flux spectra with model spectra shows that Balmer line profile shapes, and therefore the temperature structure in the line formation region, are best represented under the mixing length theory by any combination of a low mixing-length parameter $\alpha$ and a low convective structure parameter $y$.  A slightly lower effective temperature is obtained for the sun than the accepted value, which we attribute to errors in models and line opacities.
The programme stars span temperatures from 4800 to 7100~K and include a small number of population II stars.  Effective temperatures have been derived using a quantitative fitting method with a detailed error analysis.  Our temperatures find good agreement with those from the Infrared Flux Method (IRFM) near solar metallicity but show differences at low metallicity where the two available IRFM determinations themselves are in disagreement.
Comparison with recent temperature determinations using Balmer lines by Fuhrmann~(\cite{fuhrmann98, fuhrmann00}), who employed a different description of the wing absorption due to self-broadening, does not show the large differences predicted by Barklem~et~al.~(\cite{bpo:hyd}).  In fact, perhaps fortuitously, reasonable agreement is found near solar metallicity, while we find significantly cooler temperatures for low metallicity stars of around solar temperature.
}

\maketitle

\section{Introduction}

Hydrogen is by far the most abundant species in typical stellar atmospheres.  In late-type atmospheres, a small fraction of hydrogen atoms capture free electrons forming H$^{-}$ ions, which dominate the continuum opacity.  Hydrogen itself is the main opacity source for earlier spectral types. The few lines that its simple atomic structure makes in the spectrum have a very distinct sensitivity to the atmospheric properties compared to metal lines. In optical stellar spectra, absorption lines of the Balmer ($n$=2) series are commonly used to study photospheres.  The well populated lowest levels of the atom produce considerable opacity at the centre of the lines, and interactions with charged ions, electrons and other hydrogen atoms result in extended wings in high-density atmospheres.  In late-type dwarfs, these wings are believed to form very close to LTE, in the deepest photospheric layers. As most protons are bound to electrons forming hydrogen, and hydrogen influences the main continuum opacity source, changes in the hydrogen abundance are hardly reflected in the lines' strengths, and the strengths are much more weakly affected by gravity and metal abundances than perturbations to the temperature. These properties attracted the attention of stellar spectroscopists, making Balmer lines a key feature in stellar classification schemes (e.g. Morgan et~al.~\cite{morgan}). 

Detailed analyses of Balmer lines in late-type stars are more recent (e.g. Gehren~\cite{gehren1}; Fuhrmann et~al.~\cite{fuhrmann93,fuhrmann94}; van't Veer-Menneret et~al.~\cite{vbk}; Gardiner et~al.~\cite{gardiner}). These modern studies exploit progress in theory and experiment on line broadening to infer stellar effective  temperatures from Balmer lines.   Vidal~et~al.~(\cite{vcs:theory,vcs:tables}) developed a successful {\it unified theory} to model the interaction of hydrogen atoms with charged particles.  Those calculations have been recently superseded by Stehl\'e (\cite{stehle94}) and Stehl\'e \& Hutcheon~(\cite{stehle99}), who have computed Stark broadened line profiles including ion dynamic effects under the {\it model microfield method}.  A further important broadening contributor is the collisions with neighbouring hydrogen atoms. Ali \& Griem~(\cite{ali_griem:errata}) used the multipole expansion of the resonance interaction potential in the impact approximation to calculate line-widths from this process. Barklem et al. (\cite{bpo:let,bpo:hyd}, hereafter paper I and II respectively) have presented a self-broadening theory accounting also for dispersive-inductive interactions and without use of the multipole expansion. It was shown that the new description of the self-broadening would have a large impact on the computed Balmer line profiles, in particular when applied to derive effective temperatures for metal-poor dwarf stars. 

Assuming a proper understanding of the line broadening of the hydrogen lines, the most notable difficulty for the use of their wings as a temperature indicator is the fact that they are formed in very deep layers. The thermal structure of the deepest optically transparent layers in late-type stars is significantly affected by convection.  Simple modelling of surface convection is still a challenge. The commonly used mixing-length formalism incorporates unphysical parameters which are hard to connect with quantities that can be derived from observations or hydrodynamical simulations and, therefore, are difficult to constrain. This obstacle makes flux-constant homogeneous models particularly uncertain in these layers. Other theories that dispense with the free parameters in the mixing-length theory (MLT) have also been proposed (e.g. Canuto et al. \cite{canuto1}; Canuto \& Mazzitelli \cite{canuto2}); however, Gardiner et al. (\cite{gardiner}) found that after adjusting the mixing-length parameter $\alpha$, MLT  performed similarly. Semi-empirical modelling (e.g. Allende Prieto et al.~\cite{inversion}) does not offer a viable solution, as the employed metal lines do not probe layers as deep as those where hydrogen lines are formed. These theoretical problems related to establishing a model atmosphere combine with observational constraints. Balmer lines require high dispersion observations with  a large spectral coverage, and a predictable instrumental response, making possible a methodical and accurate continuum normalisation. In this situation, use of Balmer lines as a part of spectroscopic analyses requires a careful assessment of all possible sources of error.

In this paper we investigate the impact of the aforementioned line broadening theory advances in the framework of 1D model atmospheres, and attempt to identify those areas where improvement is most desirable.  In Sect.~2 we describe our observations and reduction procedure.  In Sect.~3 we describe how model spectra are computed, and put the atmospheric models in context with others.  In Sect.~4 we introduce a method for quantitative comparison of observations and model spectra, and subsequently an automated fitting procedure for deriving effective temperatures. In Sect.~5 we make a detailed survey of possible errors and their effect on effective temperatures.  Application to the solar spectrum and the programme stars is then presented.  Finally in Sect.~6 the results are compared with other work, and in Sect.~7 our conclusions are presented. 

\section{Observations and Reduction}
\label{sect:obs}

The majority of the targets were observed at the 2.5~m Isaac Newton Telescope (INT) at La Palma with the MUSICOS cross-dispersed echelle spectrograph (e.g. Baudrand \& B\"ohm~\cite{musicos}) fed by a fibre from the Cassegrain focus. Targets were selected in order to obtain a sample of dwarf and subgiant stars with a spread of temperature and metallicity, with emphasis on well studied stars.  Due to the constraints of the telescope used, only a relatively small number of metal-poor targets could be observed.  The data consist of observations taken in May 1999 and January 2000, using 2048$\times$2048 and 1024$\times$1024 pixel CCD arrays respectively. The spectra have a resolution of approximately $R\equiv \lambda/\delta \lambda \approx 30000$, and signal-to-noise ratio (SNR) per pixel of typically better than 100 at H$\alpha$ and H$\beta$, reaching 300 for some of the brighter targets.  The H$\alpha$ and H$\beta$ spectra have an average SNR of 180 and 150 respectively.   The 1999 observations were obtained as a backup programme during periods of high cirrus.  We do not expect the subsequent scattering to be significant and no attempt to remove the water vapour lines is made.   

With this spectrograph, the Balmer lines are well centred in the orders; however, the broad wings of these lines can span several orders, making determination of the continuum level a difficult process. This may be resolved by merging several consecutive orders, but a good normalisation is needed.  Flat-fielding of MUSICOS spectra taken at the INT has long been a problem, as the internal flat-field lamps available at the Cassegrain focus were designed for instruments with low and medium spectral resolution.  The MUSICOS spectrograph is a high-resolution instrument, and the front end of the fibre is relatively small compared to the entrance slits of the lower resolution instruments, thus missing a significant portion of the light from the lamps. The flat-field frames obtained during our observations were not satisfactory, seen for example by the fact that the behaviour of the continuum of orders when divided by the flat-field frames required a high-order polynomial to fit them, rather than a low-order polynomial typical of the residual difference from the stellar flux distribution and the flat-field lamp. 

\begin{figure*}
\begin{center}
\figpsaa{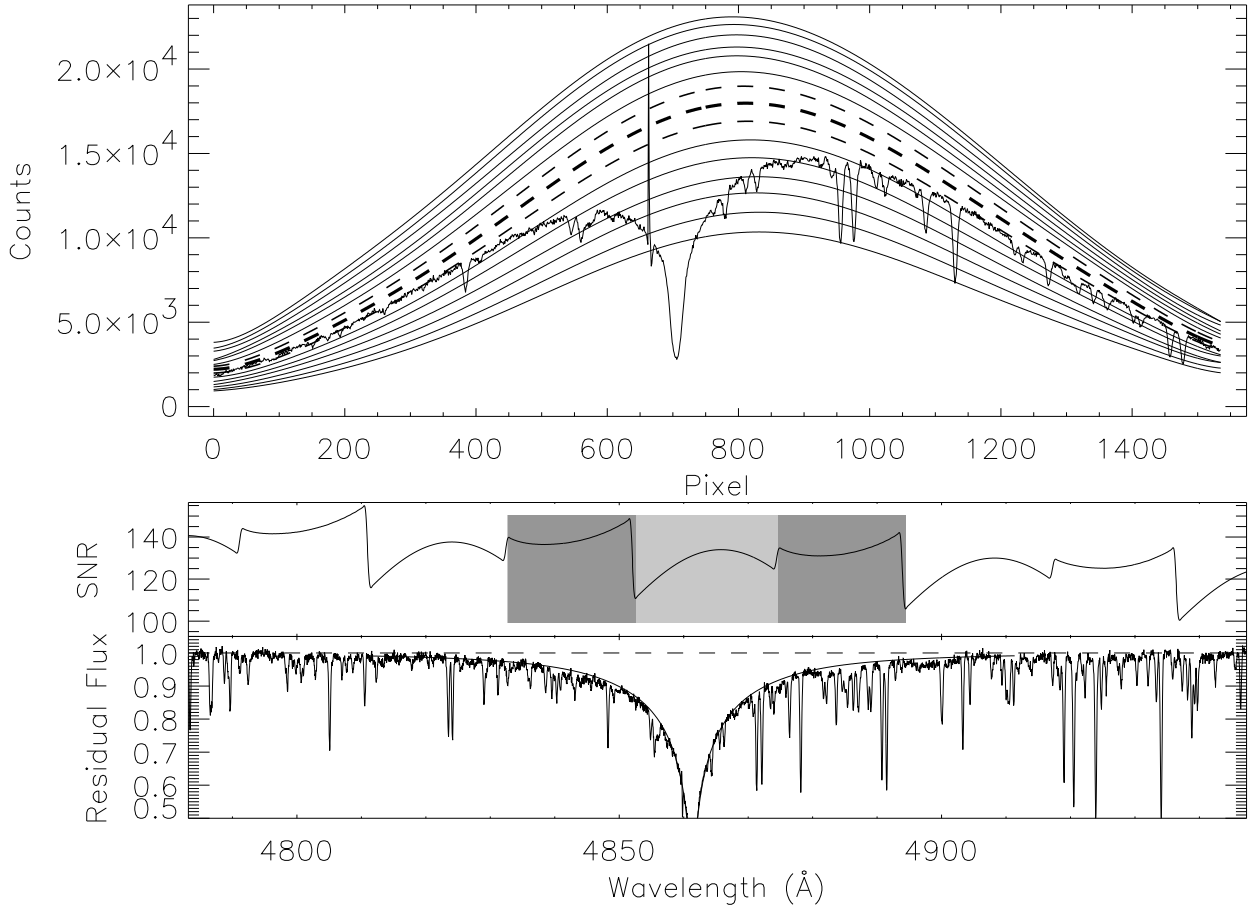}
\end{center}
\caption{An example of the continuum determination procedure, for the H$\beta$ line in HR~5447, obtained on 29 May 1999 with INT and MUSICOS.  The upper panel shows the extracted order which is centred on H$\beta$.  The smooth lines show the fit to the continuum for nearby orders where full lines are determined from the spectra, and dashed lines were determined by interpolation from the full lines.  The thicker dashed line is the continuum corresponding to the plotted spectral order.  Note the pixel number is arbitrary as the ends of orders have been removed to avoid poor behaviour of the polynomial fits.  The rapid variation of the blaze function across an order is a characteristic of the Littrow configuration used in MUSICOS.  The lower panels show the final spectrum (with best fit model spectrum) along with an estimate of the SNR per pixel \emph{in the continuum} which also serves to show the position of each order.  The upwardly convex regions are the centres of the orders, and the concave parts the overlapping ends of the orders which have higher formal SNR due to contributions from the two overlapping orders. For clarity the region corresponding to the order centred on the Balmer line (i.e. that spectrum shown in the top panel) is shaded, with the areas of overlap with the adjacent orders shaded darker. Note the cosmic ray hit in the upper panel has been removed.}
\label{fig:reduction}
\end{figure*}

For broad lines the pixel-to-pixel variations in the CCD response function are not as important as for narrow lines and so instead of using traditional flat-fields, we constructed artificial flat-fields which neglect these variations but reconstruct the blaze shape from stellar exposures. The relation between the blaze shapes of the different orders is in principle a smoothly changing function, and because in general only a few orders contain broad spectral lines the blaze shape for these orders can be determined by interpolation. The spectral layout of frames taken with the MUSICOS spectrograph is very well adapted to this approach, because not only is the number of orders in the frame large (of order 50), but also the coverage in the dispersion direction is large (typically 60--100\AA), tracing the shape of the blaze function far out in its wings.  Orders have significant overlap in wavelength, even at H$\alpha$ where the overlap is smallest, the amount depending on the CCD array used. To determine the shape of the blaze function we used the fact that for our data set the extracted profiles contain only absorption lines. We then performed a polynomial fit to the upper points in each extracted order. This method fails if there are no continuum points available due to the presence of a broad hydrogen line in the order.  A surface fit of the neighbouring orders is then used to determine the artificial flat-field at any problematic orders.  The regions for interpolation are selected interactively, which may include other problematic regions such as strong metal lines or molecular bands.  This procedure can also be used if the frames are first divided by a traditional flat-field exposure, as the flat-field blaze-function is still smoothly changing.  On the larger CCD fringing at H$\alpha$ made this procedure necessary for the rectification of this line.  However, for other cases we did not divide the exposures as we found the variation of the orders was in general smoother.  Orders are extracted and scattered light corrected using standard IRAF {\it echelle} package routines, before normalisation as described above using IDL. Finally the wavelength calibration is determined, orders are merged together weighted by the SNR at the wavelength in the order, and wavelengths are corrected to the the stellar rest frame.  The method is illustrated in Fig.~\ref{fig:reduction} for one of the more difficult cases (low SNR$\approx$120 and broad line).

For Procyon, in addition to spectra from INT/MUSICOS, a very high quality spectrum has been obtained by Allende~Prieto~et~al.~(\cite{allende_procyon}) with the McDonald Observatory 2.7~m telescope and the {\it 2dcoud\'e} spectrograph (Tull et~al.~\cite{tull}) for a large spectral region including H$\beta$. The spectra have SNR of about 1000 at H$\beta$, and a resolution of approximately $2 \times 10^5$ (see Allende~Prieto~et~al.~\cite{allende_procyon} for details).  The same continuum placement techniques described for INT/MUSICOS were applied to these observations.  

The spectrum of HD 103095 (Gmb 1830) was acquired on 28 and 29 April 1990 with the McDonald 2.7~m telescope and the Coud\'e spectrograph. The observations employed a conventional grating in first order and a colour filter to eliminate light from higher orders. The detector was a 800$\times$800 pixel CCD. The dispersion is about 0.12~\AA/pixel and the SNR was a few hundred. This spectrum was processed using IRAF being normalised using a second-order polynomial fit.  

Comparison of reduced spectra from different exposures of the same targets, revealed similar results to those of Fuhrmann et~al.~(\cite{fuhrmann93}), namely an internal consistency of approximately $1\%$ of the continuum flux, dependent on SNR.  Most of this error probably stems from the determination of the continuum placement.  The error can be reduced through co-adding a number of independent observations (Fuhrmann et~al.~\cite{fuhrmann93}). Direct comparison of moon spectra with the FTS atlas (Kurucz~et~al.~\cite{kurucz:atlas}) degraded to appropriate resolution shows agreement at the level of $0.5$--$1\%$, although small differences in wavelength calibration make the direct comparison difficult.  Comparison of the McDonald and INT Procyon spectra shows similar agreement, despite quite different instrumental setups.  We will test this consistency later by examining scatter in temperatures derived from different observations.

\section{Models and Synthetic Spectra}
\label{sect:models_and_spectra}

In this work we employ 1D LTE plane-parallel model atmospheres from the MARCS code (Asplund~et~al.~\cite{osmarcs} version, see this reference and those therein for details).  The adopted version of MARCS employs opacity sampling techniques, except in the infrared where opacity distribution functions are deemed sufficiently accurate.  This circumvents problems of opacity distribution functions namely allowing free choice of individual metal abundances (cf. Fuhrmann et~al.~\cite{fuhrmann97}). Convection is modelled under MLT which is parameterised in terms of a number of parameters, the most important of which for model atmospheres and our discussions are the mixing-length $l$, which is usually expressed in units of the pressure scale height $H_p$ as the mixing-length parameter $\alpha \equiv l/ H_p$, and the structure parameter $y$ describing the temperature structure within convective elements (see Henyey~et~al.~\cite{henyey} for details).   A $\log\tau_\mathrm{ross}$ ($\tau_\mathrm{ross}$ is Rosseland mean optical depth) grid suitable for computation of Balmer lines was chosen.  The depth grid extends in to $\log\tau_\mathrm{ross}=2$ and the grid is finest in the region where these lines are formed $-1 < \log\tau_\mathrm{ross} < 0.6$.  

Models are computed with scaled-solar photospheric abundances from Anders \&\ Grevesse~(\cite{anders_grevesse}) except for and most importantly the iron abundance where a ``low'' value of $\log(N_{\mathrm{Fe}}/N_{\mathrm{H}})+12 = 7.50$ is adopted in agreement with the meteoritic value (e.g. Grevesse \& Sauval~\cite{gs98}).  A microturbulence of 1.5~km~s$^{-1}$ was used in all cases for computing the models. The MLT parameters used were $\alpha=1.5$, $y=0.076$ the default MARCS setting, and $\alpha=0.5$, $y=0.5$ following Fuhrmann~et~al.~(\cite{fuhrmann93}).  Hereafter we will use MARCS to refer to the former and MARCS05 to the latter.  The reasons for these choices are discussed later. Each grid was also recomputed with enhancement of the alpha-elements by 0.4 dex, which we will denote in discussions with an ``a'' (e.g. MARCS05a).  When referring specifically to the grids with scaled-solar abundances we will use an ``s''.  The model grids are computed with spacings of 100~K in $T_\mathrm{eff}$, 0.1 dex in $\log g$ and 0.5 dex in [Fe/H], within which we interpolate specific models.

Synthetic model flux spectra are computed as described in paper II. In short, spectra are computed assuming LTE using SYNTH (Piskunov~\cite{piskunov:synth}), Stark broadening is described by the model-microfield method calculations of Stehl\'e \&\ Hutcheon~(\cite{stehle99}) and self-broadening calculations of paper II are used.  Radiative broadening is included, as is an estimate of the helium collisional broadening (based on rescaling of Barklem \&\ O'Mara~\cite{bo:pd}), although both effects are small.  We will refer to this as the STEHLE+BPO broadening recipe.  For comparison with previous work we also do calculations substituting the Ali \& Griem~(\cite{ali_griem:errata}) resonance broadening theory for the self-broadening calculations, everything else remaining the same.  This will be referred to as the STEHLE+AG broadening recipe. In our implementation of the Ali \& Griem theory we only consider the broadening of the 2p state, since the 2p--$n$d transition dominates (see discussion in paper II).  Note that this differs from Fuhrmann~et~al.~(\cite{fuhrmann93}) where both the 2p and $n$p state are included.

\subsection{Comparison to other models}
\label{sect:compmodels}

To put the MARCS05 models which will be used in most of this work in context, we now compare them differentially with some other models.  To compare the models particularly as regards computed Balmer line profiles, we made comparisons by finding the $T_\mathrm{eff}$'s of the MARCS05 Balmer line profiles which best match profiles produced by other models using least-squares minimisation.  This should indicate approximately the difference in $T_\mathrm{eff}$ that would be found from the input model relative to MARCS05 for the given line.  STEHLE+BPO recipe was used to produce all profiles and only the wings are considered.  

First we compared with other common solar models, both semi-empirical and Kurucz~(\cite{kurucz:cdroms}) theoretical flux-constant models.  Table~\ref{tab:modeldiffs_sun} shows the difference in temperature of the MARCS05 model needed to match the input profile (i.e.~$T_\mathrm{eff}$(best match)$-5777$) for the first two Balmer lines.  Large differences with semi-empirical models are seen, which produce much weaker Balmer lines than the MARCS05 solar model.  When comparing with solar observations the semi-empirical models produce lines too weak and the MARCS05 too strong when STEHLE+BPO is employed (see paper II and Sect.~\ref{sect:sun}). The differences with Kurucz models in Table~\ref{tab:modeldiffs_sun} stem from different MLT parameters and inclusion of overshoot in the KOVER model.  We should point out that after publication of paper II it was realised that the MARCS model used in that paper was in fact computed with $\alpha=1.25$ and $y=0.5$, the same as the Kurucz models, not $\alpha=1.5$ and $y=3/(4\pi^2)$ as stated, hence the reasonable agreement of the profiles with KNOVER.  One should note that despite the rough agreement of H$\beta$ temperatures from KNOVER and MARCS05, the line shapes are markedly different, which was true of H$\beta$ in all cases.

\begin{table}
\begin{center}
\caption{Differences between other solar models, in terms of effective temperatures from Balmer lines, of our MARCS05 models.}
\label{tab:modeldiffs_sun}
\begin{tabular}{crr}
\hline
Model   &     $\Delta T_{\mathrm{eff}}$(H$\alpha$)  &     $\Delta T_{\mathrm{eff}}$(H$\beta$)  \\
        & (K) & (K) \\
\hline
GS                              &   $-81$  &  $-196$   \\
(Grevesse \&\ Sauval~\cite{gs99}) &&\\
HM         &  $-114$  &  $-192$   \\ 
(Holweger \&\ M\"uller~\cite{holmul})  &&\\
MISS       &  $-132$  &  $-250$   \\
(Allende~Prieto~et~al.~\cite{inversion}) &&\\
KOVER  &   $-24$  &  $-214$   \\
(Kurucz~\cite{kurucz:cdroms}, see paper II) &&\\
KNOVER  &   $+73$  &   $-10$   \\
(Kurucz~\cite{kurucz:cdroms}, see paper II) &&\\
\hline
\end{tabular}
\end{center}
\end{table}

\begin{figure}
\figpsa{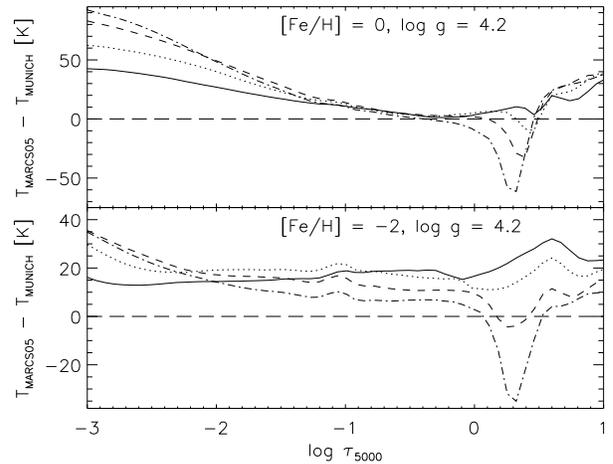}
\caption{Differences between T--$\tau$ structures of MARCS05 and MUNICH models, for $T_\mathrm{eff}=5200$, 5600, 6000, and 6400~K (full, dotted, dashed and dot-dashed lines respectively), all for $\log g=4.2$.  The upper and lower panels show [Fe/H]$=0.0$ and [Fe/H]$=-2.0$ cases respectively.  }
\label{fig:munich_marcs_ttau}
\end{figure}

The differences with the Munich group models (see~Fuhrmann~et~al.~\cite{fuhrmann97}, hereafter MUNICH) are of interest to understand the discrepancies between the $T_\mathrm{eff}$ values from this work and Fuhrmann~(\cite{fuhrmann98,fuhrmann00}).  Comparisons were made for a range of stellar temperatures and metallicities and the results are shown in Table~\ref{tab:modeldiffs_mafags}.  Alpha-enhancement is included for all models below solar metallicity.  A single variation of gravity is also shown.  We see that the effect on H$\alpha$ is generally larger and can be as high as 40~K for hotter stars.  For most cases the corrections are positive, which indicates a hotter temperature would be found using MARCS05 than MUNICH models.  Fig.~\ref{fig:munich_marcs_ttau} compares temperature structures from both model grids.  We see that in the upper regions of the atmospheres, MARCS05 models typically have slightly hotter and less steep T--$\tau$ structures.  We note that for the cooler models, particularly at low metallicity, the temperature is hotter throughout the entire atmosphere.

\begin{table}
\begin{center}
\caption{Differences between MARCS05 and MUNICH models in terms of effective temperatures derived from Balmer lines.  }
\label{tab:modeldiffs_mafags}
\begin{tabular}{crr}
\hline
Model   &     $\Delta T_{\mathrm{eff}}$(H$\alpha$)  &     $\Delta T_{\mathrm{eff}}$(H$\beta$)  \\
$T_{\mathrm{eff}}/\log g/$[Fe/H]        & (K) & (K) \\
\hline
5200/4.2/0.0         &    $+6$  &   $+3$   \\
5600/4.2/0.0         &    $+9$  &  $+11$   \\
6000/4.2/0.0         &   $+21$  &  $+13$   \\
6400/4.2/0.0         &   $+39$  &   $+1$   \\
&&\\
5200/4.6/0.0         &    $+3$  &   $-2$   \\
5600/4.6/0.0         &    $+6$  &   $+1$   \\
6000/4.6/0.0         &   $+16$  &   $-1$   \\
6400/4.6/0.0         &   $+29$  &   $-6$   \\
&&\\
5200/4.2/$-$1.0      &    $-8$  &  $-15$   \\
5600/4.2/$-$1.0      &    $-1$  &  $-11$   \\
6000/4.2/$-$1.0      &   $+14$  &   $-7$   \\
6400/4.2/$-$1.0      &   $+28$  &   $-6$   \\
&&\\
5200/4.2/$-$2.0      &    $+2$  &   $-8$   \\
5600/4.2/$-$2.0      &    $+2$  &   $+1$   \\
6000/4.2/$-$2.0      &   $+11$  &   $+1$   \\
6400/4.2/$-$2.0      &   $+31$  &   $+2$   \\
\hline
\end{tabular}
\end{center}
\end{table}

\section{Fitting Method}
\label{sect:method}

Quantifying the comparison between a synthetic profile and an observed profile, allows us to automate the fitting process through minimisation of the chosen statistic, and also provides a statistical measure of the goodness-of-fit.  This further allows the subjectivity associated with such fitting, usually done by inspection, to be shifted as we will demonstrate.  We employ a reduced $\chi^2$ statistic, namely
\begin{equation}
\chi^2 = \frac{1}{N-M} \sum^N_{i=1} \left( \frac{f_i - F_i}{\sigma_i} \right)^2
\end{equation}
where $N$ is the number of wavelength points, $M$ is the number of free parameters (here one, namely $T_{\mathrm{eff}}$), $f_i$ is the synthetic residual flux, $F_i$ the observed residual flux, and $\sigma_i=1/\mathrm{SNR}$.  For the value of $\sigma_i=1/\mathrm{SNR}$ we estimated a constant average value of the SNR (in the continuum) for the given Balmer line observation.   Despite the fact that the formal SNR of our observations varies across the line as shown in Fig.~\ref{fig:reduction}, the regions with highest SNR are those at the end of the orders where continuum placement may be less certain.  Therefore these should not be given a higher weight, and a constant average value is more appropriate. 

For our case of Balmer lines, blending lines must be accounted for.  Calculations including the metal lines are impractical as current spectral line databases have not yet reached a level of completeness or accuracy where all observed lines in the solar spectrum in the Balmer line regions are well accounted for (see Fig.~9 in paper II).  Furthermore, such an approach would require an iterative procedure with more free parameters.  

A much simpler approach is to attempt the identification of spectral windows across the line which are expected to be free from blends in the stars of interest.  This was done by using the high resolution FTS solar spectrum (Kurucz~et~al.~\cite{kurucz:atlas}) and the best of our spectra of hotter and cooler stars of around solar metallicity (namely Procyon and HR 8832) to identify windows which would be free from obvious blending across most of our sample.  High resolution spectra of a metal-rich K dwarf would be particularly useful in this respect, as this perhaps represents the hardest case, but to our knowledge such spectra are not available.   We make the assumption that unblended spectral windows do in fact exist and that we see them in the above mentioned high quality spectra.  This is reasonable for the sun and Procyon at least at H$\alpha$ and H$\beta$ based on examination of the spectra, but may for other stellar parameters or higher Balmer lines be questionable. Once these windows had been identified we considered other desirable qualities of the windows \emph{as a set}. Firstly the Balmer line cores, which are formed high in the atmosphere in non-LTE are excluded, only the pressure broadened wings are considered.  We attempted to give approximately equal weight to all parts of the line wings, despite a natural tendency for the inner-wings to show less blends due to saturation by the hydrogen line.   We note that if the windows are evenly distributed across the line wings, and a least-squares fitting is employed such as in this work, the procedure is quite similar to matching the equivalent width of the wings of the line, and the information on the line shape is not biased towards any part of the line.  Furthermore, a reasonable number of windows need to be employed so as to avoid the possibility that statistical fluctuations in the observations affect the fitting procedure.  One should also be sure that windows are appropriate for the spectral resolution of the observations, and this will be discussed in more detail shortly.

When comparing model spectra with observations in this way, we employ the same spectral windows, or a slightly reduced subset, for \emph{all stars}.  Rejection of some windows is necessary in specific cases, such as the fast-rotating early F stars where metal lines are very broad and encroach on these unblended regions, or if an atmospheric line by chance falls within a window in the reference frame of a given star, or if a window happens to fall in the line core.  This method restricts the human interaction in the fitting process, once the initial windows are decided, to simply this choice of rejecting some windows.  By following this approach we attempt to avoid the situation where one must distinguish between noise fluctuations in the spectra and possible weak metal lines, which can be particularly difficult in metal-poor stars with lower SNR observations.  Thus in essence, the subjectivity involved in the usual fitting by visual inspection is shifted to this choice of windows.

In cool stars, where self-broadening is important, one must be cognizant that both descriptions of this process used in this work employ the impact approximation. The \emph{extreme} limit of validity for the self-broadening in the STEHLE+BPO recipe is given by
\begin{equation}
\Delta \lambda_{\mathrm{max}} = 3.654 \times 10^{-7} \lambda^2 \frac{T^{(2+\alpha)/4} \;0.145^\alpha}{\sqrt{\sigma_{10^4}}}
\end{equation}
where the result $\Delta \lambda_{\mathrm{max}}$ is in \AA, with $\lambda$ the central wavelength in \AA, $T$ the temperature in K, $\sigma_{10^4}$ the cross-section in atomic units for $v=10^4 \mathrm{m\;s}^{-1}$ and $\alpha$ the dimensionless velocity parameter (not to be confused with the MLT $\alpha$ parameter).  Typical values for the sun are discussed in paper II.  The region of validity varies through the stellar atmosphere, but based on the assumption that the line wings are predominantly formed in the region of the atmosphere around optical depth unity, a reasonable estimate for a given stellar profile can be obtained by setting $T=T_{\mathrm{eff}}$. Spectral windows falling outside this region of validity are rejected.  For consistency, even when the STEHLE+AG recipe is used we use the same limit as for STEHLE+BPO, although a calculation based on the Ali \& Griem~(\cite{ali_griem:errata}) cross-section would give a slightly larger region of validity.

\begin{figure*}[p]
\begin{center}
\figpsax{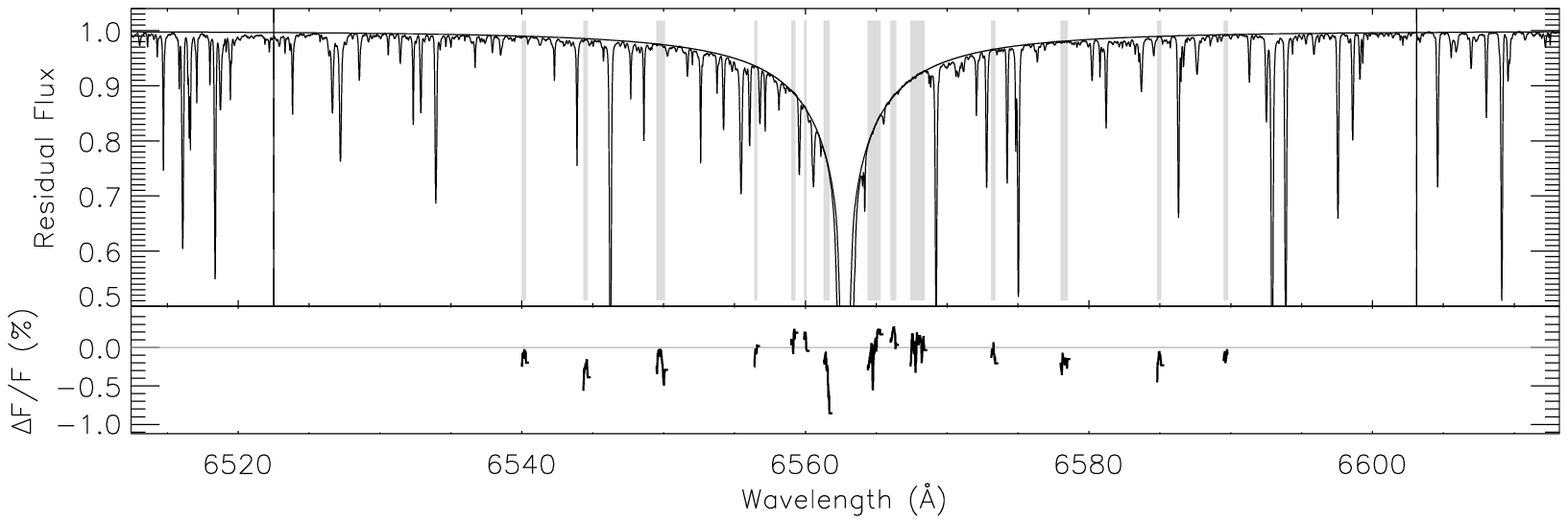}
\figpsax{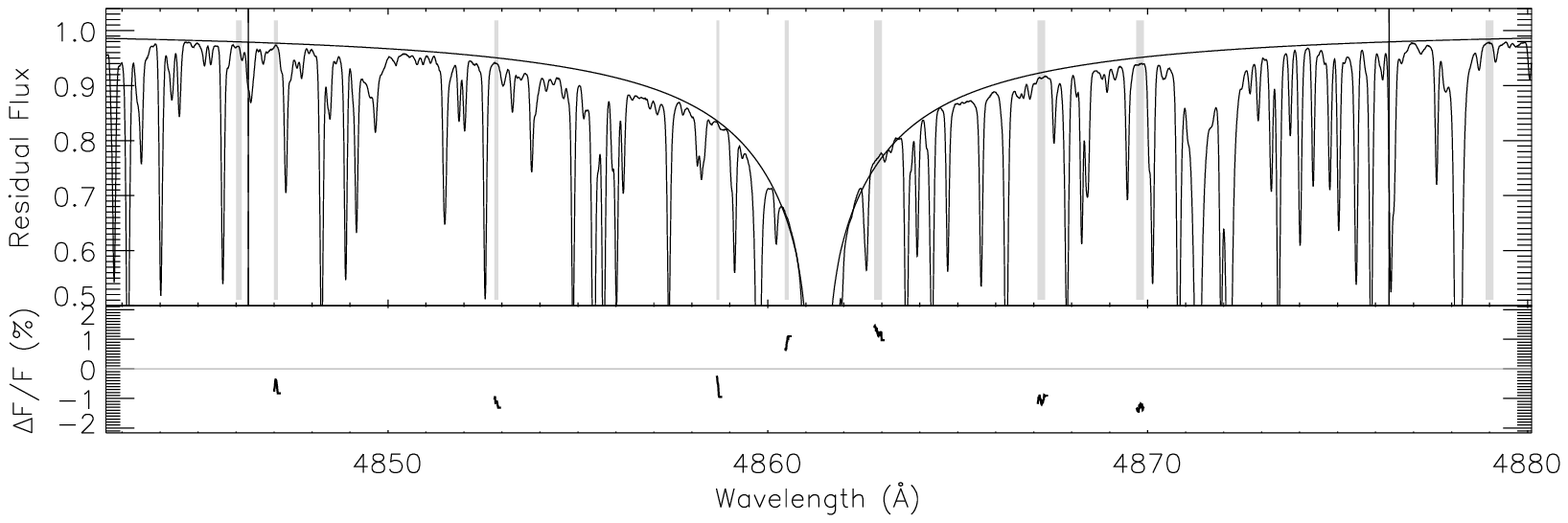}
\end{center}
\caption{The fitting method for solar H$\alpha$ and H$\beta$ profiles, here showing the best fit found with MARCS05 and STEHLE+BPO which corresponds to the parameters in Table~\ref{tab:stars}.  The shaded regions show the windows used for determining the $\chi^2$ statistic.  The full vertical lines show the estimated limit of validity of the impact approximation.  Note the windows outside this region are rejected and thus no residual is plotted. }
\label{fig:fit_sun}
\end{figure*}

\begin{figure*}[p]
\begin{center}
\figpsax{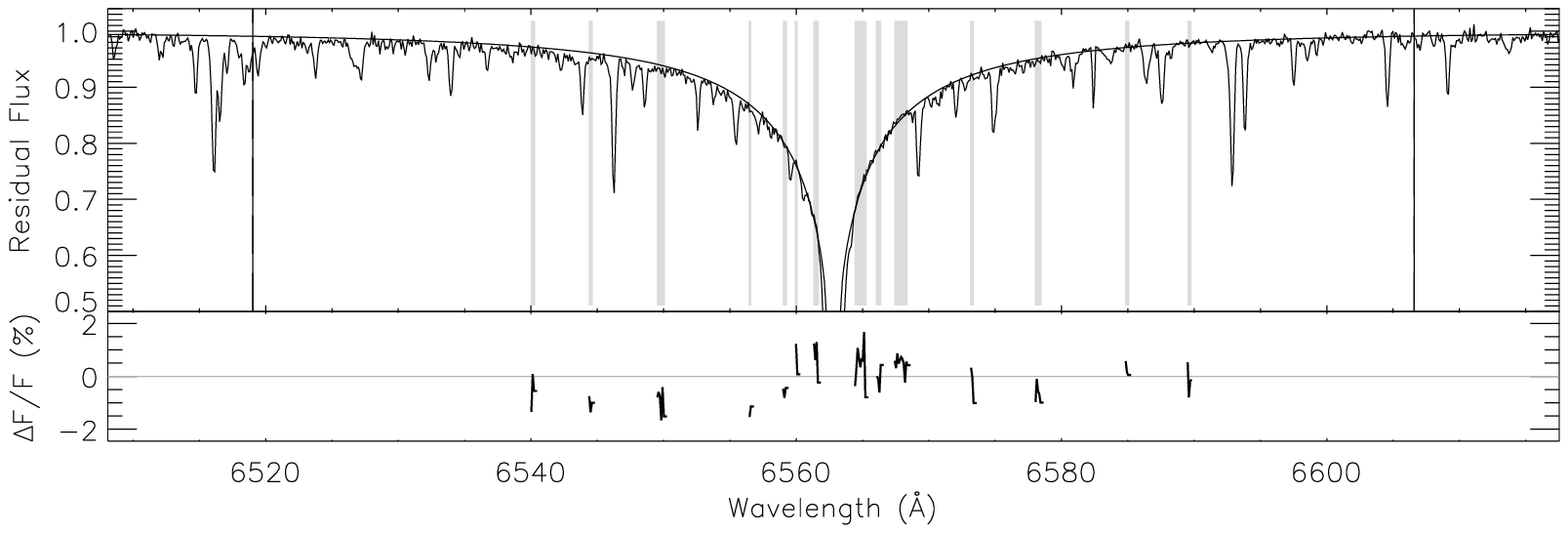}
\figpsax{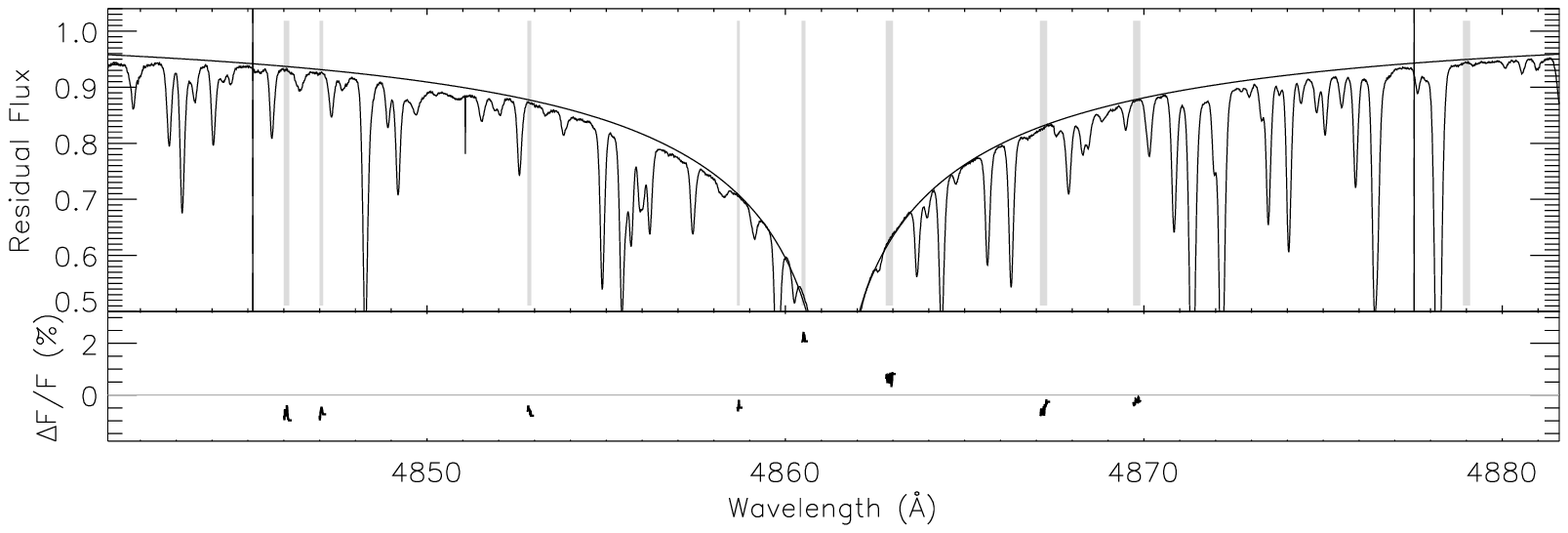}
\end{center}
\caption{As in Fig.~\ref{fig:fit_sun} for Procyon H$\alpha$ and H$\beta$ profiles.}
\label{fig:fit_proc}
\end{figure*}

If errors in the data are normally distributed, minimisation of the $\chi^2$ statistic provides the maximum likelihood estimate of the model parameters, in our case $T_{\mathrm{eff}}$.  Our errors will in fact be Poissonian, approaching a normal distribution at high SNR.  A more robust (insensitive to departures from normal distribution) statistic to better account for outliers might be of benefit.

For each star we computed a grid of H$\alpha$ and H$\beta$ profiles for a given model grid at 10~K intervals in $T_\mathrm{eff}$ around an initial guess based on literature values. The $\chi^2$ statistic was always found to be well behaved varying smoothly with $T_\mathrm{eff}$, approximately parabolic in form with only a single minimum (no other local minima).  We therefore searched this grid for the best fitting profile, and finally performed a polynomial fit to the $\chi^2$ statistic values near the minimum and determined the predicted minimum in order to improve the resolution of the fitting to 1~K, while saving computing time, a procedure which was verified by direct calculations.  Figs.~\ref{fig:fit_sun} and~\ref{fig:fit_proc} show the best fits for the sun and Procyon from this method, including our final chosen spectral windows.  Despite obtaining observations of H$\gamma$ of reasonable quality for some stars, this line is not considered in this work due to the problem of choosing suitable windows and the limited validity of the self-broadening calculations.

In using such a method there is a natural tendency that a chosen window may not be entirely free from blends, as wings of nearby lines may encroach on the apparent window or unseen weak blends exist. One must therefore expect that the fitting method may very slightly over-estimate $T_{\mathrm{eff}}$ in a systematic manner.  This is likely of all such fitting, automated or not.

\subsection{Influence of spectral resolution}
\label{sect:resolution}

It is important to understand the effect of spectral resolution on our results.  Sufficient resolution is needed such that a suitable number of windows between blending lines are available.  Once again, this is true not just of our fitting method but any approach.  There must exist a limiting resolution for which all windows will be affected by the instrumental broadening of metal lines, but above which there is at least one unaffected window, and this limit differs with star, and spectral region.  For our purposes, a single small spectral window is insufficient, since as discussed above, at the minimum, we would like to have several spectral windows evenly distributed across the Balmer line.
  
Ideally, windows should be chosen such that for the test cases the derived effective temperature does not differ between the high resolution spectra and the result obtained with the same spectra degraded to the resolution of our observations $R\approx 30000$, or at least introduces only a small error (say $<$10~K). This requirement was quite easily fulfilled for H$\alpha$ in the sun.  However, for H$\beta$ this was problematic.  It proved difficult to find a set of windows which fulfilled both our criteria of a $T_\mathrm{eff}$ in reasonable agreement with the high resolution result, and a reasonable coverage of the whole line profile shape.  The windows eventually used for H$\beta$ were chosen as a compromise between both criteria, which resulted in a temperature from the degraded spectra of 17~K hotter than the result using high resolution spectra in both stars.  This error is considered acceptable, in view of gaining some information on the fit of the line shape, though we expect this error to increase in cooler stars.   Because of this, and quite significant extra blending, H$\beta$ will not be employed for the coolest stars in our sample.

The results for different resolutions using the eventual set of spectral windows are incorporated in Table~\ref{tab:tests}.  A brief survey of different resolutions indicates that a spectral resolution of at least $R\approx50000$ is more appropriate for gaining both $T_\mathrm{eff}$ and information on the line profile shape from H$\beta$ for solar-type stars.    

\begin{table*}
\begin{center}
\caption{Derived $T_\mathrm{eff}$ values for the sun and Procyon using different or degraded observations.  Results are for MARCS05s models.  Note all $R$ and SNR values are approximate.}
\label{tab:tests}
\begin{tabular}{cccccc}
\hline
Observation & $R$ &  SNR &  SNR            & $T_\mathrm{eff}$ H$\alpha$ &  $T_\mathrm{eff}$ H$\beta$ \\
            &     & H$\alpha$ &  H$\beta$  &   (K)                      &      (K) \\
\hline
\multicolumn{6}{l}{\underline{Sun}} \\
Kitt Peak FTS atlas & $>3\times 10^5$& 3000 & 3000  &  5733     &   5723  \\
---                      & 50000 & 3000   &  3000   &  5735     &   5729  \\
---                      & 30000 & 3000   &  3000   &  5739     &   5740  \\
INT/MUSICOS (co-added)   & 30000 & 375    &  240    &  5743     &   5748  \\
 --- (single exposure)   & 30000 & 170    &  110    &  5721     &   5711  \\
&&&&&\\
\multicolumn{6}{l}{\underline{Procyon}} \\
McDonald/2dcoud\'e       & $2\times 10^5$ &---& 1000&  ---      &  6474   \\
---                      & 30000 & ---    &  1000   &  ---      &  6481   \\
INT/MUSICOS (co-added)   & 30000 & 250    &  160    &  6538     &  6487   \\
 --- (single exposure)   & 30000 & 120    &  80     &  6498     &  6532   \\
\hline
\end{tabular}
\end{center}
\end{table*}

\section{Results}

\subsection{Error Estimates}
\label{sect:errors}

Due to the complicated interplay of several mechanisms in the formation of hydrogen Balmer lines, estimating the sources of error and their behaviour with stellar parameters is similarly complicated.  Fuhrmann~et~al. (\cite{fuhrmann93,fuhrmann94}) discuss many of these effects.  Here, we have taken a quantitative approach to the errors.  By making a reasonable guess of the error in a number of possible sources, we have surveyed how these errors translate into errors in $T_\mathrm{eff}$ through $\chi^2$ comparisons of model profiles.  For consistency, the same windows as will be used for the observations are used, and test calculations employing the whole line profile indicate this does not introduce much error, which is also a confirmation that our windows do not introduce significant bias to certain parts of the line profile.  Our error estimates are summarised in Table~\ref{tab:errors}.  Test calculations showed that these errors \emph{do vary} with MLT parameters.  For example, for the $T_\mathrm{eff}$=6000~K and [Fe/H]=$-2$ case, Table~\ref{tab:errors} indicates H$\alpha$ to be quite sensitive to gravity while H$\beta$ is rather insensitive. Test calculations for this case with MARCSa models in place of MARCS05a find the sensitivity to gravity of H$\alpha$ and H$\beta$ to be roughly the same, a 0.1 dex change corresponding to about $\pm34$~K for both lines.  Therefore, our results are somewhat dependent on our choice of MLT parameters.

Differences between the often used Vidal~et~al.~(\cite{vcs:tables}, hereafter VCS) and Stehl\'e~(\cite{stehle94}) calculations perhaps give an indication of the magnitude of the error in the Stark broadening calculations, and these differences are reported in Table~\ref{tab:errors}.  Test calculations introducing a 5$\%$ error to the Stark broadening found results of similar magnitudes.  VCS calculations always resulted in a stronger profile and therefore a cooler temperature.  Stehl\'e's~(\cite{stehle94}) calculations for proton perturbers are known to have reasonable agreement with experiment, better than VCS where ion dynamics are neglected. However, one should consider that while protons dominate in hot stars, in cooler stars the perturbing ions progressively become more a mix of protons and heavier ionised nuclei. Self-broadening for hydrogen is less well studied than Stark broadening, no experimental results currently existing.  Based on error estimates for hydrogen broadening of metal lines using similar theoretical techniques we estimate the error at around 5$\%$ (e.g. Barklem \& O'Mara~\cite{bo:jpb}), though we note that the interaction is fundamentally different due to the resonance interaction.  As the broadening by helium is only estimated here we adopt a rather large error bar of 50$\%$.  Typical errors in the gravity and metallicity are estimated at 0.1 dex.  Possible interdependency of errors makes their correct combination non-trivial, thus we chose a somewhat {\it ad hoc} procedure. The errors discussed to this point are grouped into two categories, broadening and stellar parameters, and these two errors are combined in quadrature supposing they are independent and random.  The totals for each category are found by simply summing.  These fixed errors are sub-totalled in Table~\ref{tab:errors}.  To estimate errors for a given line we interpolate in these fixed values which are then combined with the error sources discussed below which vary from case to case.

As discussed in Sect.~\ref{sect:obs} observational errors are estimated at approximately 1$\%$, and so the estimated errors in Table~\ref{tab:errors} corresponding to a 0.5$\%$ shift in the continuum placement represent perhaps a best case which we assume for our best spectra.  In computing errors we interpolate in these values, and scale the result depending on the quality (SNR) of the observations employed. We assume 1.5$\%$ for our lowest quality spectra.  In Sect.~\ref{sect:sun} we will see for our adopted model grid there is a discrepancy of order 50--60~K in the solar $T_\mathrm{eff}$ value with the known value.  Assuming negligible error in observations, gravity and metallicity, approximately half of this can be explained by our estimates of the error due to broadening theory given in Table~\ref{tab:errors}.  The remainder of the discrepancy can most likely be attributed to the models.  Thus, for the solar case we estimate a conservative model error of order 40~K.  It must be anticipated that this error will vary with stellar parameters, but it is difficult to estimate how.  The error due to possible variation in MLT parameters is estimated from the difference between $T_\mathrm{eff}$ values derived with MARCS and MARCS05 models, which is added to this 40~K.  This above procedure amounts to using the sun to approximately calibrate our error bars.  The effect of deviation from the solar or alpha-enhanced abundance patterns (i.e. abundance scatter) is estimated by half the difference between the results obtained from MARCS05s and MARCS05a model grids.  We also include a fitting error, including resolution effects, of 20~K and 40~K for H$\alpha$ and H$\beta$ respectively, conservative values based on the tests we performed in Sect.~\ref{sect:resolution} (see Table~\ref{tab:tests}).  Totals are shown in Table~\ref{tab:errors} for the case where there is no error due to convection or abundance scatter, and the best case observations.  Our final errors will account for these additional factors on a case by case basis.

\begin{table*}
\tabcolsep=1.5mm
\begin{center}
\caption{Estimated errors in $T_\mathrm{eff}$ (in K) corresponding to introduced errors in the input physics, stellar parameters or observations for MARCS05 models, with MARCS05a models used for [Fe/H]$=-1$ and $-2$ cases. Results are given for various $T_\mathrm{eff}$ at both solar and low metallicity (quoted as $T_\mathrm{eff}/$[Fe/H]).  All cases use $\log g = 4.2$, except the sun where $\log g =4.44$.  Errors are introduced (amounts discussed in the text) and the corresponding changes in derived $T_\mathrm{eff}$ observed.  Signs show the direction of the change, for the corresponding change in the input physics or stellar parameter. Errors are usually symmetric, but if not the largest is quoted.  We emphasise these errors vary with chosen MLT parameters.  For a given stellar spectrum the continuum error is interpolated from the listed values and scaled depending on the SNR, and the values of $\Delta T_\mathrm{eff}$(convection) and $\Delta T_\mathrm{eff}$(abundance scatter) are computed case by case and given in Table~\ref{tab:stars}.}
\label{tab:errors}
\scriptsize
\begin{tabular}{ccccccccccc}
\hline
Introduced     &  Sun   &   5000/0.0   &   6000/0.0    &   7000/0.0   &  5000/$-$1.0   &   6000/$-$1.0  &   7000/$-$1.0  &  5000/$-$2.0   &   6000/$-$2.0  &   7000/$-$2.0   \\
error                   & H$\alpha$/H$\beta$    & H$\alpha$/H$\beta$    & H$\alpha$/H$\beta$    & H$\alpha$/H$\beta$    & H$\alpha$/H$\beta$    & H$\alpha$/H$\beta$    & H$\alpha$/H$\beta$ & H$\alpha$/H$\beta$    & H$\alpha$/H$\beta$    & H$\alpha$/H$\beta$    \\
\hline
Stark-broadening  &&&&&&&&&&\\
 VCS                          &  $-$5/$-$15     & $-$3/$-$10     & $-$7/$-$18       & $-$19/$-$28      & $-$1/$-$7       & $-$7/$-$17      & $-$18/$-$26     & 0/$-$3          &  $-$8/$-$17       & $-$17/$-$25    \\
Self-broadening   &&&&&&&&&&\\ 
    $\pm5\%$                   & $\mp$14/$\mp$7  & $\mp$10/$\mp$6 & $\mp$13/$\mp$6   & $\mp$8/$\mp$3    & $\mp$15/$\mp$13 & $\mp$23/$\mp$11 & $\mp$11/$\mp$3  & $\mp$23/$\mp$20  &  $\mp$38/$\mp$15 & $\mp$12/$\mp$4 \\
He-broadening     &&&&&&&&&&\\
 $\pm50\%$                     & $\mp$6 /$\mp$4  & $\mp$5/$\mp$3  & $\mp$6/$\mp$3    & $\mp$4/$\mp$2    & $\mp$6/$\mp$7   & $\mp$10/$\mp$6  & $\mp$5/$\mp$2   & $\mp$10/$\mp$11  &  $\mp$17/$\mp$8  & $\mp$6/$\mp$2  \\ 
$\log g$  &&&&&&&&&&\\
 $\pm0.1$ dex                  &      ---  & $\mp$6/$\mp$6  & $\mp$9/$\mp$3    & $\mp$3/$\pm$8    & $\mp$13/$\mp$12 & $\mp$25/$\mp$7  & $\mp$3/$\pm$11  & $\mp$33/$\mp$17  &  $\mp$52/$\mp$11 & $\mp$5/$\pm$11 \\
$[$Fe/H$]$  &&&&&&&&&&\\
 $\pm0.1$ dex                  &      ---  & $\pm$19/$-$2   & $\pm$7/$\mp$17   & $\mp$25/$\mp$31  & $\pm$35/$\pm$20 & $\pm$17/$\mp$7  & $\mp$14/$\mp$15  & $\pm$16/$\mp$7   &  $\pm$6/$\mp$4   & $\pm$6/$\pm$6  \\
\hline
&&&&&&&&&&\\
Sub-total                      &  $\pm$25/$\pm$26 & $\pm$31/$\pm$21 & $\pm$31/$\pm$34 & $\pm$42/$\pm$51 & $\pm$53/$\pm$42 & $\pm$58/$\pm$37 & $\pm$38/$\pm$40 & $\pm$59/$\pm$48 &  $\pm$86/$\pm$43 & $\pm$37/$\pm$35 \\
\hline
Continuum    &&&&&&&&&&\\
  $\pm0.5\%$                   &  $\pm$38/$\pm$29 & $\pm$36/$\pm$23 & $\pm$29/$\pm$19 & $\pm$34/$\pm$21 & $\pm$33/$\pm$25 & $\pm$38/$\pm$23 & $\pm$35/$\pm$21 & $\pm$38/$\pm$25 &  $\pm$56/$\pm$26 & $\pm$36/$\pm$21 \\
&&&&&&&&&&\\
Model                 & \multicolumn{10}{c}{$\longleftarrow$ 40 + $|\Delta T_\mathrm{eff}$(convection)$|$ $\longrightarrow$} \\
&&&&&&&&&&\\
Abundance scatter & \multicolumn{10}{c}{$\longleftarrow$ $|\Delta T_\mathrm{eff}$(abundance scatter)$|$ $\longrightarrow$} \\
&&&&&&&&&&\\
Fitting \& resolution & \multicolumn{10}{c}{$\longleftarrow$ 20/40 $\longrightarrow$} \\
\hline
&&&&&&&&&&\\
Total (best case)  &  $\pm$69/$\pm$72 & $\pm$65/$\pm$65 & $\pm$61/$\pm$69 & $\pm$70/$\pm$79 & $\pm$76/$\pm$75 & $\pm$82/$\pm$71 & $\pm$69/$\pm$73 & $\pm$83/$\pm$79 &  $\pm$111/$\pm$76 & $\pm$68/$\pm$70 \\
\hline
\end{tabular}
\end{center}
\end{table*}

\subsection{The sun}
\label{sect:sun}

The sun is the most important test case since high quality observations are available and the stellar parameters are well known.  We follow Fuhrmann~et~al.'s~(\cite{fuhrmann93}) approach and consider the MLT parameters as free and attempt to calibrate them at least approximately using the solar case.  We say approximately since the appropriate convective efficiency for representing Balmer lines must be expected to vary across the HR diagram (e.g. Ludwig et al.~\cite{ludwig}). Both Fuhrmann et~al.~(\cite{fuhrmann93}) using solar observations, and Steffen \& Ludwig~(\cite{steffen}) using hydrodynamical simulations of the sun, have found that a low value of $\alpha\approx0.5$ is needed to represent Balmer lines, both with $y$ set at 0.5.  We note that Steffen \& Ludwig~(\cite{steffen}) have shown using 2D hydrodynamical models that, at least for flux-constant MLT models, ``different mean [temperature] stratifications are needed to represent different spectroscopic properties of an inhomogeneous stellar photosphere'' and therefore we should not expect the convective efficiency from observations of H$\alpha$ and H$\beta$ which will be investigated here to be the same as that required to represent other observational quantities, even other Balmer lines.  

\begin{figure*}
\begin{tabular}{cc}
\figpsab{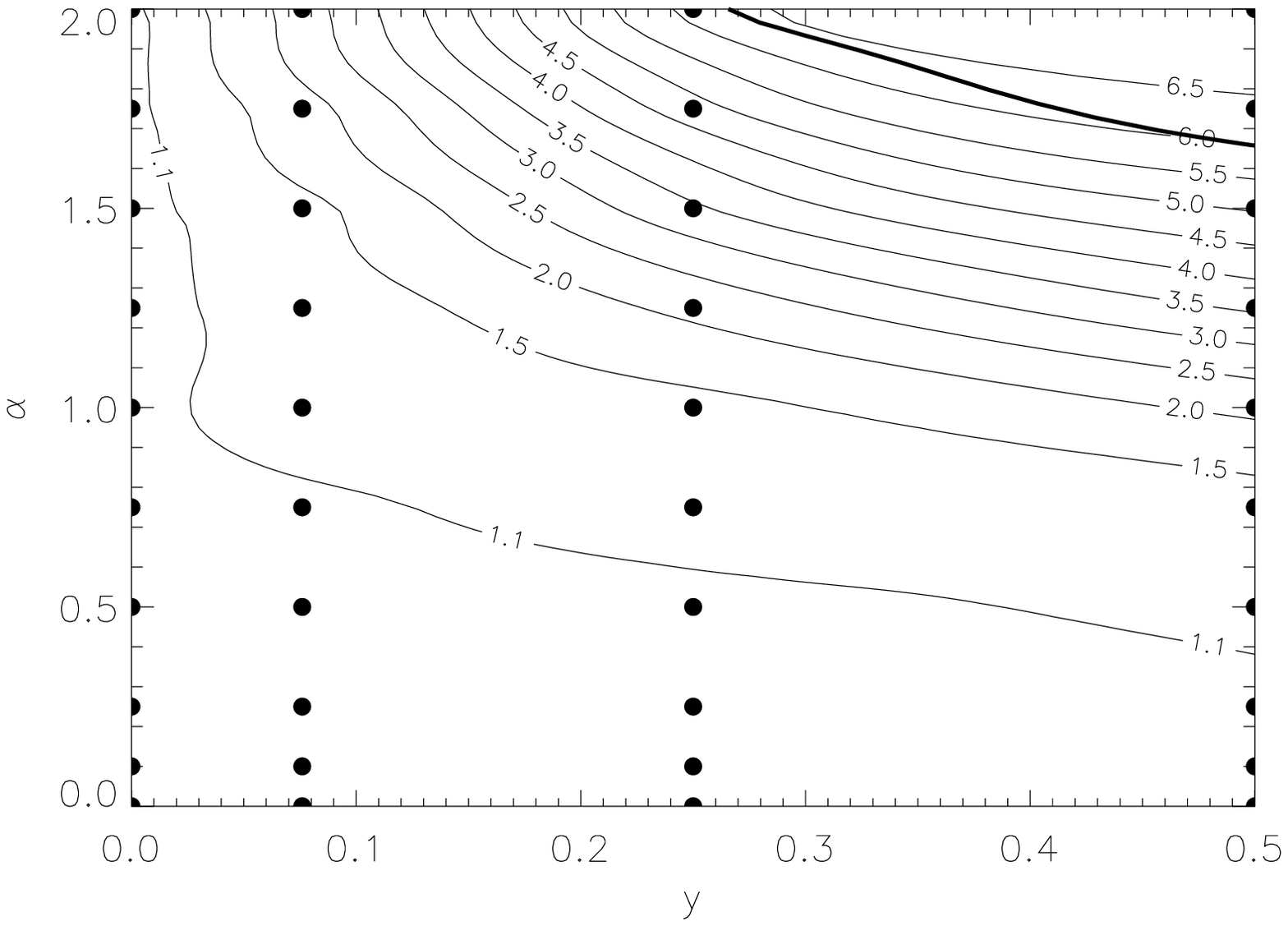} & \figpsab{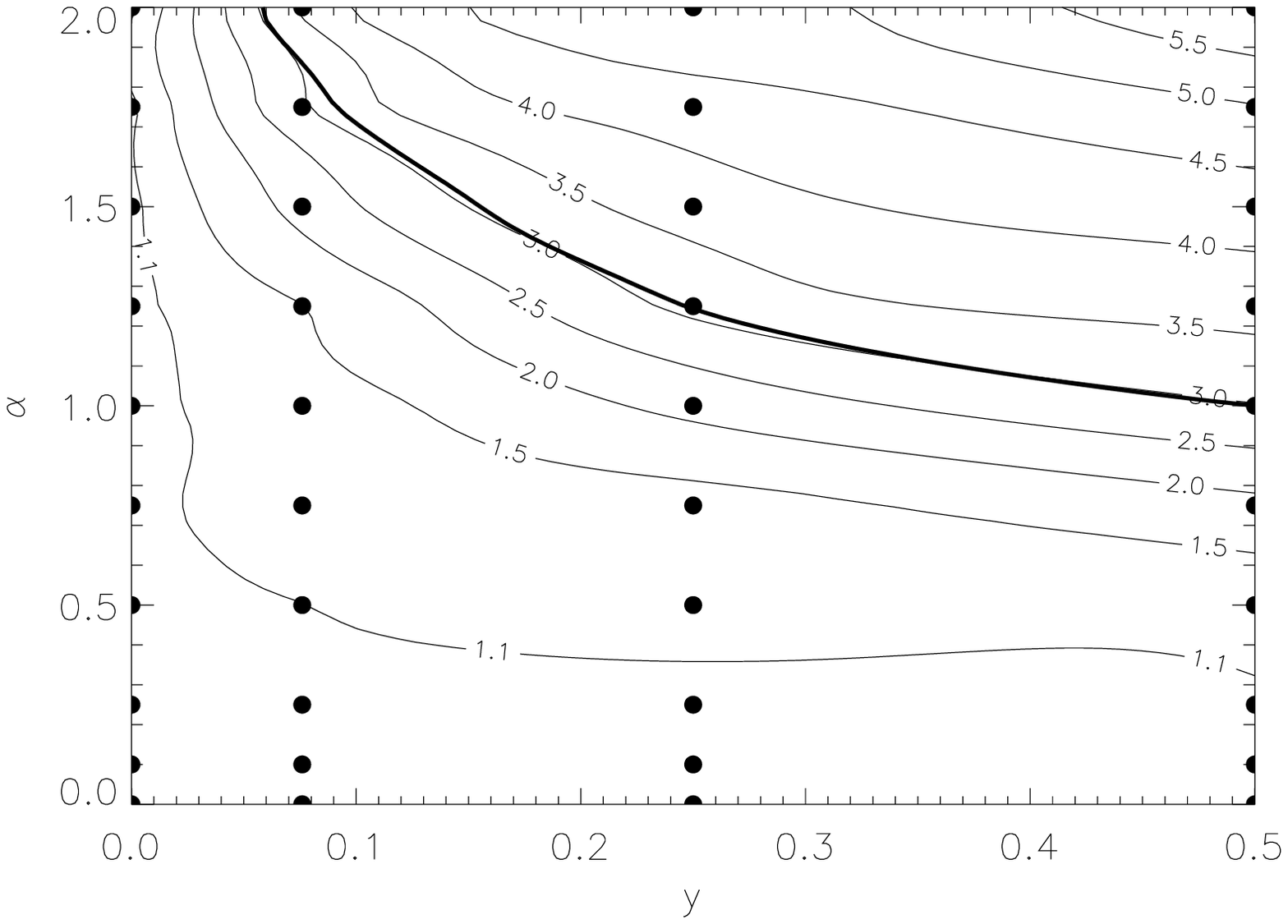} \\
\figpsab{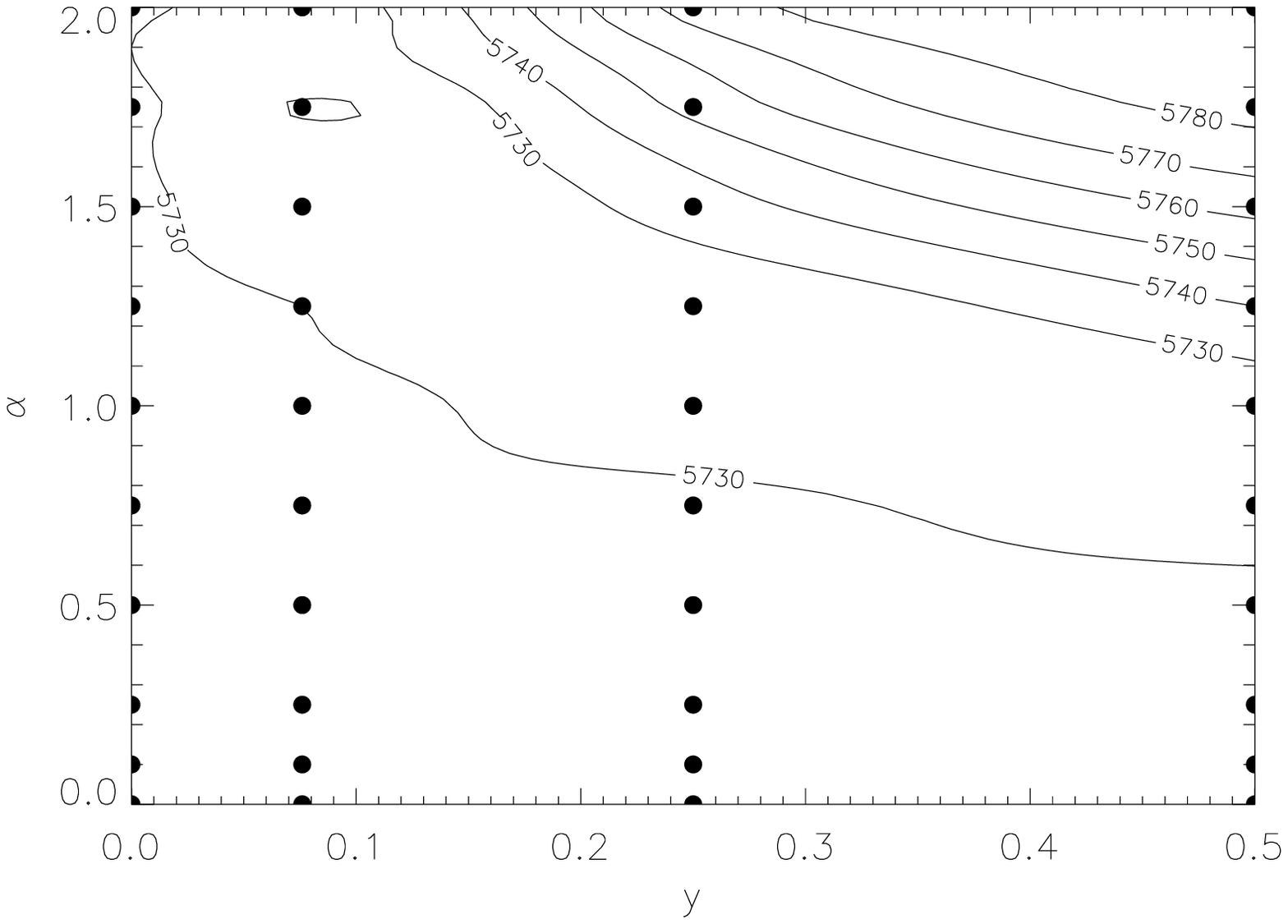} & \figpsab{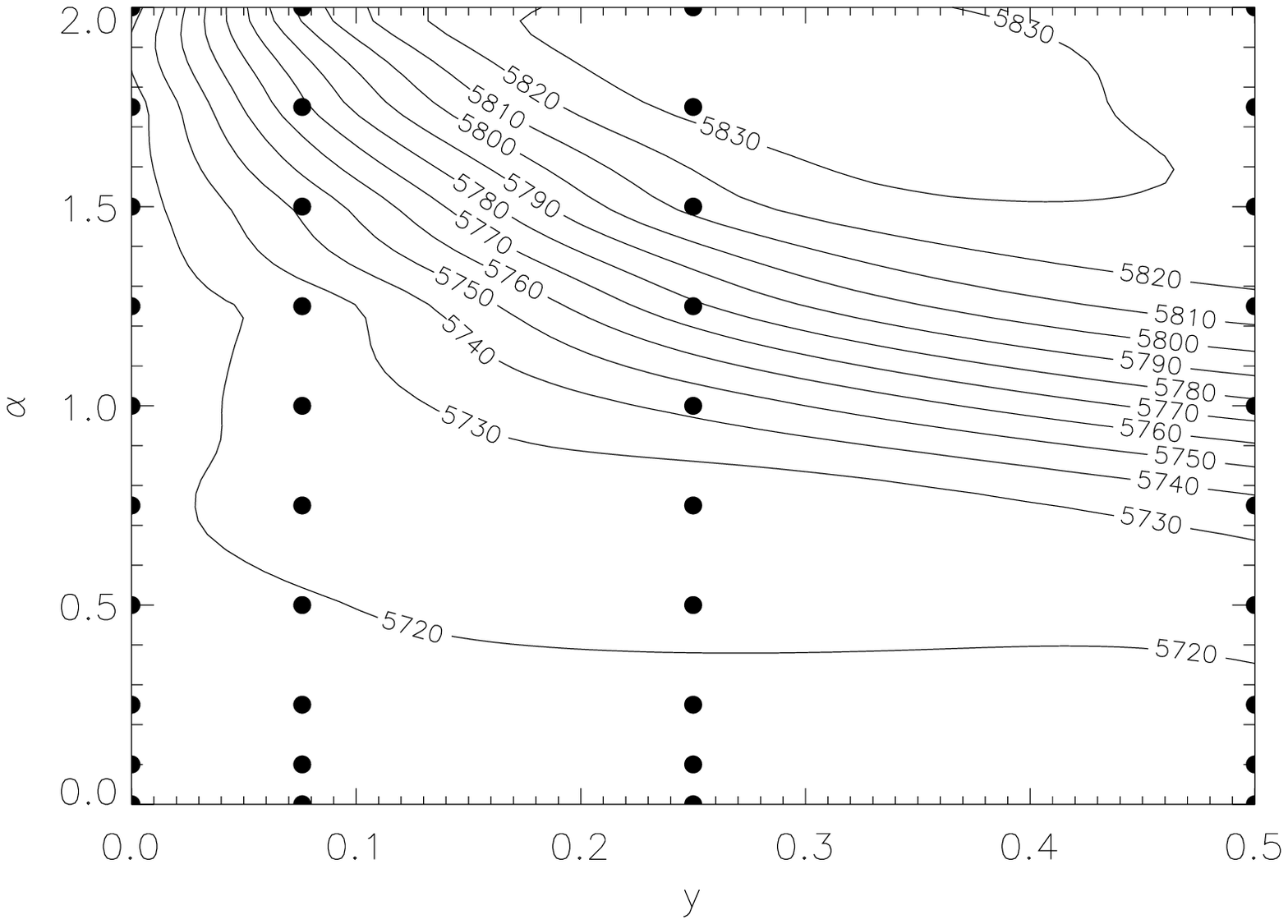} \\
\figpsab{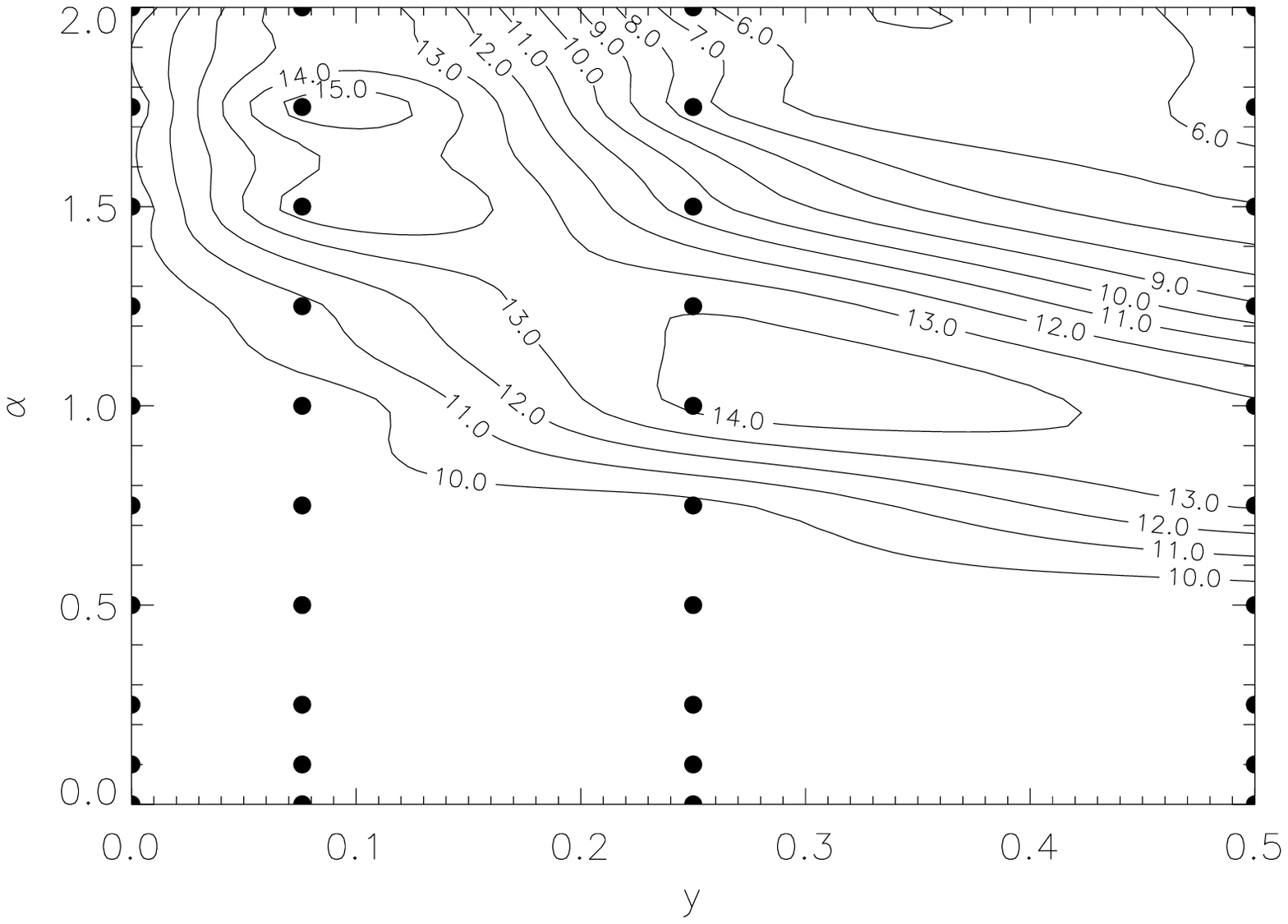} & \figpsab{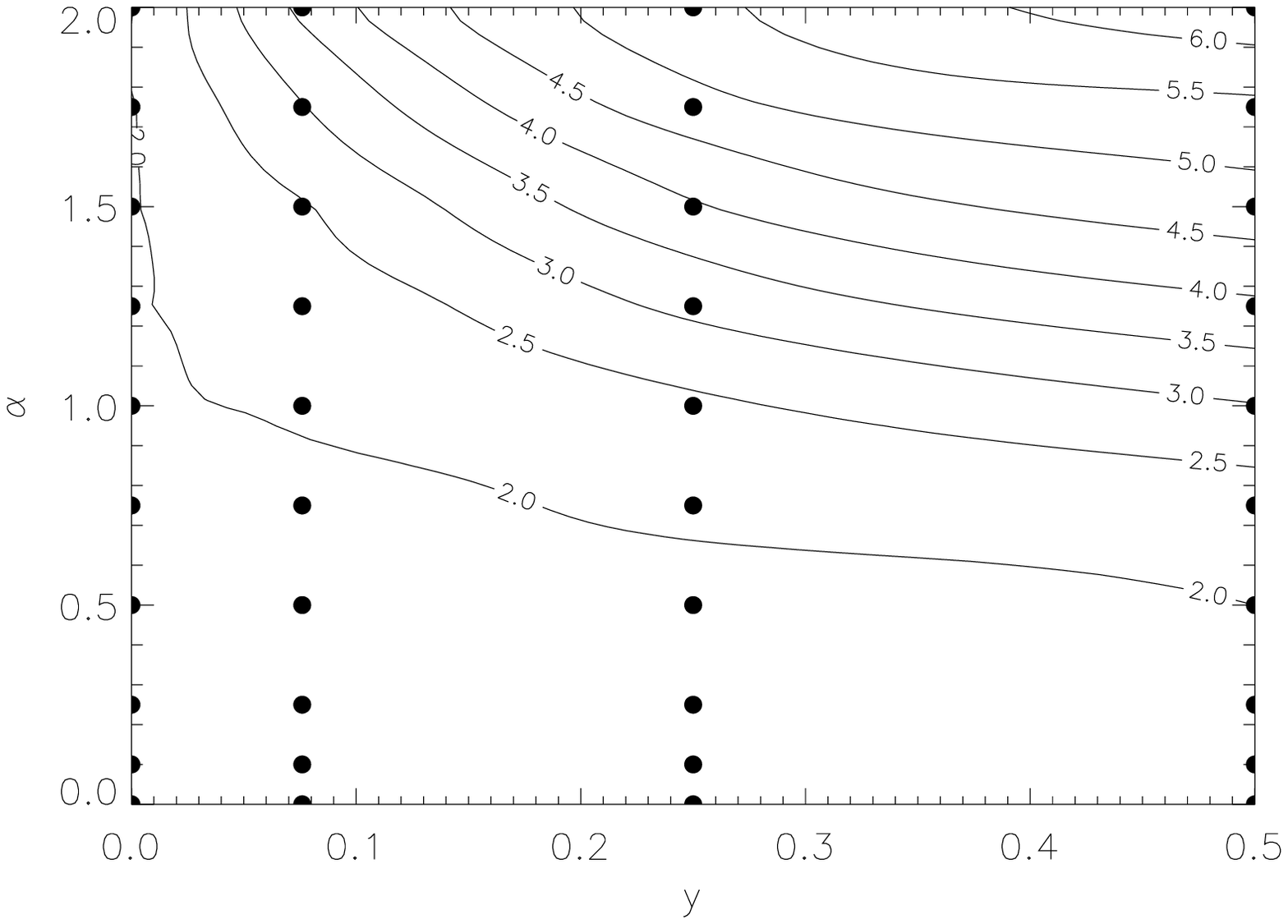} \\
\end{tabular}
\caption{Survey of Balmer line fits for the observed solar flux spectra with different MLT parameters using STEHLE+BPO recipe.  The left column of plots refers to H$\alpha$ and the right to H$\beta$. In each case the uppermost plot is the $\chi^2$ values for the best fit in the case where $T_\mathrm{eff}$ is considered a free parameter, the thick line on these plots indicating the locus where $T_\mathrm{eff} = 5777$~K is found.  The middle plot shows the corresponding behaviour of the best fit $T_\mathrm{eff}$ values. The lowermost plots show $\chi^2$ for the case where $T_\mathrm{eff}$ is fixed at 5777~K.  All $\chi^2$ are normalised so the best fit for a given line is 1.0, (i.e. both top and bottom plots of each column are on the same scale) which in both cases corresponds to the plateau in the uppermost plots at low $\alpha$ and $y$.  Filled circles show the points actually computed and the contours are found by interpolation (minimum curvature surface) and thus any structure on scales smaller than the computed grid should be ignored. }
\label{fig:chisq_sun}
\end{figure*}

This work differs from Fuhrmann et~al.'s~(\cite{fuhrmann93}) in that, in addition to the mixing-length parameter $\alpha$, the convective structure parameter $y$ is allowed to vary, and the STEHLE+BPO broadening recipe is employed (see Fuhrmann et al.~\cite{fuhrmann93} for details of their broadening recipe). We have investigated the solar case with the FTS solar atlas~(Kurucz~et~al.~\cite{kurucz:atlas}) using a number of models where the MLT parameters $\alpha$ and $y$ have been varied within reasonable limits.  Two approaches are taken.  Firstly, we derive $T_\mathrm{eff}$ considering it a free parameter for a given MLT $\alpha$, $y$ set and record the $\chi^2$ value.  Secondly, we set the effective temperature at 5777~K and survey the $\chi^2$ value across the different MLT parameters.  The results are displayed in Fig.~\ref{fig:chisq_sun}, where all $\chi^2$ values are normalised so the best possible fit for a given line has $\chi^2=1$.  Steps in $\chi^2$ of 0.5, from 1 to 20, are plotted (except in one case where we use steps of 1 for clarity).  Differences below this level correspond to only small differences in the line profiles and are probably meaningless.  The contour at 1.1, where it exists, is also shown to illustrate the plateau at $\chi^2=1$.

It is seen that no single parameter set shows itself to be especially preferred on the basis of agreement with line shape and $T_\mathrm{eff}$.  A simple approximate picture is that agreement with the line shape indicates correct temperature structure in the line formation region, and agreement with $T_\mathrm{eff}$ indicates correct temperature at the line formation region, although the reality is somewhat more complex.   The line profile shapes for both H$\alpha$ and H$\beta$ are best fit anywhere on the plateau at low $\alpha$ and low $y$ with a $T_\mathrm{eff}$ some 50~K lower than the accepted value.   If we force $T_\mathrm{eff} = 5777$~K the fits to line profile shapes for both H$\alpha$ and H$\beta$ are quite poor, typically being too deep.  Furthermore, a single $\alpha, y$ combination will not give the correct $T_\mathrm{eff}$ for both lines, seen by the fact that the loci of $\alpha, y$ values giving the correct solution (thick lines) do not overlap.  More generally one can see from Fig.~\ref{fig:chisq_sun} that Balmer lines do not well constrain the MLT parameters, since an increase in $\alpha$ can be compensated by a decrease in $y$, further strengthening Steffen \& Ludwig's~(\cite{steffen}) assertion that Balmer lines should not be seen as evidence for low efficiency of solar-type convection.

On the assumption that the FTS observations employed are of high accuracy in comparison with the model spectra, one interprets this as a shortcoming of either the model atmospheres or the broadening theory, though most likely a combination of both.  In paper II using the same broadening theory we found that the semi-empirical model of Holweger \& M\"uller~(\cite{holmul}), which matches limb-darkening better than MARCS models (Blackwell et~al.~\cite{blgs}), in fact gives lines too shallow.  Therefore it is certainly feasible that this discrepancy could be due to models.  Based on these results we decided to adopt the same parameters as Fuhrmann et~al.~(\cite{fuhrmann93}) namely $\alpha$=0.5 and $y$=0.5 as our default values since this parameter combination reproduces the line shapes and therefore the temperature stratification in the line forming region as well as any other choice.  As we see in Fig.~\ref{fig:fit_sun}, the fit to H$\beta$ is not perfect showing differences of order $1\%$ with a trend across the line, again indicating some deficiency in either models or line broadening.   Although a lower value of $y$ is perhaps more justified physically on the basis of the diffusion approximation (Henyey~et~al.~\cite{henyey}), adopting these parameters will make comparisons with Fuhrmann~(\cite{fuhrmann98, fuhrmann00}) more straightforward.  In any case, all calculations will be repeated with $\alpha$=1.5 and $y$=0.076 as a test of sensitivity to this choice.  The error in $T_\mathrm{eff}$ introduced by this choice of MLT parameters was discussed in Sect.~\ref{sect:errors}.

In addition to testing models, the sun is also the best test case for our observational data and reduction procedures.  By applying our fitting method to different observations we can test their accuracy.  A high SNR spectrum of the moon was produced from co-added spectra, as well a spectrum at lower SNR more typical of our other observations from a single exposure.  The two spectra were reduced completely independently.  Comparison of the derived $T_\mathrm{eff}$ values with that from the FTS atlas degraded to the same resolution presented in Table~\ref{tab:tests} shows a scatter of typically $\pm 15$~K for H$\alpha$ and $\pm 30$~K for H$\beta$.

\subsection{Procyon}

The F5 star Procyon (HR~2943) is of fundamental importance as it is one of the few dwarf stars other than the sun accessible to key direct measurements due both to its proximity and binary nature.  Allende Prieto et~al.~(\cite{allende_procyon}) find $T_\mathrm{eff} = 6530 \pm 49$~K and $\log g = 3.96 \pm 0.02$ in good agreement with other works (e.g. Di Benedetto~\cite{dibenedetto}, Fuhrmann et~al.~\cite{fuhrmann97}).  

As mentioned, we have a high quality spectrum of H$\beta$ for Procyon, and therefore this star serves as a second important test case.  Results shown in Table~\ref{tab:tests} indicate that for our window set $R\approx30000$ is adequate for H$\beta$, a result which is expected since blending is less pervasive.  When comparing spectra from different observations at $R\approx30000$, a scatter of about $\pm 50$~K is seen.  This is more than for the solar case, which we interpret as being due to the lower SNR of the single exposure leading to a less reliable continuum placement. However, we should also consider the possibility that the broader lines in Procyon make the reduction procedure slightly less intrinsically reliable.

For this star, with the best available spectrum, we find 6538$\pm$82~K from H$\alpha$ and 6474$\pm$143~K from H$\beta$. Our final adopted result (weighted average, rounded) of 6520$\pm$80~K is in good agreement with the above mentioned results.  We note also that this result agrees within error with IRFM results of 6579$\pm$100~K from Alonso et~al.~(\cite{alonso}) and 6601$\pm$75~K from Saxner \&\ Hammerb\"ack~(\cite{saxner}), but does not agree with the value found by Edvardsson et~al.~(\cite{bdp}) of 6704~K from $b-y$ colours.  We note, that as for the solar case the fit to H$\beta$ shown in Fig.~\ref{fig:fit_proc} is not perfect showing differences with a trend across the line profile.

\subsection{The sample}

Using our automated fitting procedure, effective temperatures were derived from all four model grids (MARCSs, MARCSa, MARCS05s, MARCS05a) using both broadening recipes, adopting surface gravities and metallicities from the literature.  Fitting for all four models with both broadening recipes allows some interesting comparisons, and is also useful in estimating errors as discussed in Sect.~\ref{sect:errors}.  Following the well known enhancement of alpha-elements, for stars with metallicity less than [Fe/H]$=-0.5$ the result of the alpha-element enhanced grid is used, otherwise the scaled-solar result is adopted. Unless otherwise stated, all discussion below refers to the STEHLE+BPO results.

\begin{figure*}
\figpswbig{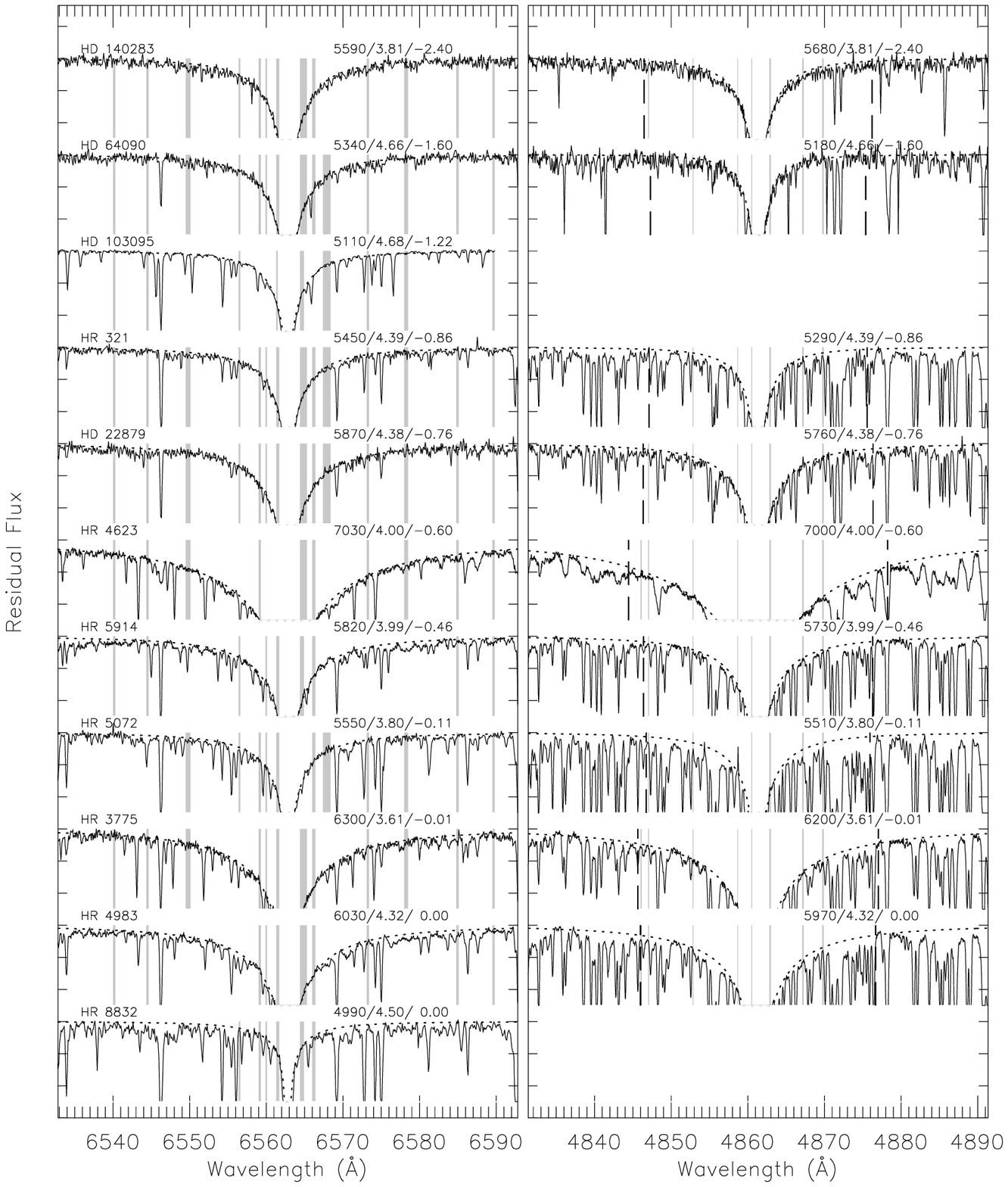}
\caption{Selected Balmer line spectra and fits.  Full lines are the observed spectra and dotted lines the synthetic spectra corresponding to our best fit with MARCS05 and STEHLE+BPO, the parameters for which are given to the upper right of each profile ($T_\mathrm{eff}/\log g/$[Fe/H]).  Tick marks on the vertical scale correspond to 0.1 in residual flux, and continua are at every 3rd tick mark.  Double-thickness dashed vertical lines show the approximate validity region of the impact approximation of the BPO theory (for H$\alpha$ this always lies off the plot and is therefore only seen for H$\beta$).  Shaded regions show windows used for fitting for each case. }
\label{fig:spectra_examples}
\end{figure*}

\begin{figure}
\figpswh{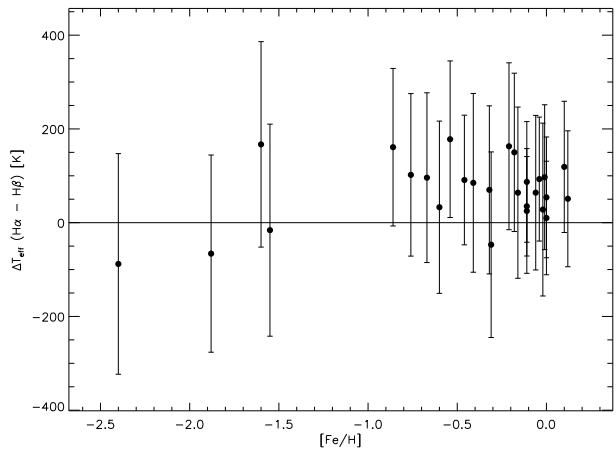}
\caption{The difference between derived $T_{\mathrm{eff}}$ from H$\alpha$ and H$\beta$ using MARCS05 and STEHLE+BPO. }
\label{fig:ha-hb}
\end{figure}

The derived values of $T_{\mathrm{eff}}$ from MARCS05 for each line are presented in Table~\ref{tab:stars}, the adopted temperature being the weighted average of the results from both lines.  Some selected example fits to the spectra are shown in Fig.~\ref{fig:spectra_examples}.  Fig.~\ref{fig:ha-hb} plots the difference between $T_{\mathrm{eff}}$ from H$\alpha$ and H$\beta$ against metallicity (no trend is seen with $T_{\mathrm{eff}}$), and suggests that H$\beta$ gives temperatures which are cooler by 69$\pm$70~K than those from H$\alpha$.  Essentially the same result is found for STEHLE+AG recipe. When MARCS models are employed we find better agreement in that H$\beta$ is typically cooler by 11$\pm$75~K, though the line shapes become noticeably worse.   Comparing the fits to H$\beta$ for MARCS and MARCS05 models it was found that on average $\chi_{\mathrm{MARCS}}^2 /\chi_{\mathrm{MARCS05}}^2 = 1.85 \pm 0.75$, all stars indicating a better fit to the shape with MARCS05.  H$\alpha$ also showed better fits for MARCS05 for most stars although the effect was not as strong, as would be expected due to the decreased sensitivity of H$\alpha$ to the convection parameters. Very similar results were obtained with the STEHLE+AG recipe.  One motivation for using a quantitative profile comparison was to look for trends with stellar parameters in the relative goodness-of-fit of model profile shapes from different convection parameters to observations.  However, we note that the $\chi_{\mathrm{MARCS}}^2 /\chi_{\mathrm{MARCS05}}^2$ ratio is expected to be dependent on SNR, which was confirmed from experiments using solar and Procyon spectra. Although the result converges for high SNR, at the typical SNR used here it was found the ratio would be significantly underestimated. Therefore care must be exercised in comparisons of such statistics between different stars and models.

\begin{table*}
\begin{minipage}{\textwidth}
\begin{center}
\caption{The programme stars and derived effective temperatures.  Sources indicates the source of adopted $\log g$ and [Fe/H] values in that order. Estimated errors due to convection and abundance scatter are shown, where $\Delta T_{\mathrm{eff}}$(conv)$=T_{\mathrm{eff}}$(MARCS)$-T_{\mathrm{eff}}$(MARCS05) and $\Delta T_{\mathrm{eff}}$(abund)$=\pm0.5\times(T_{\mathrm{eff}}$(MARCS05a)$-T_{\mathrm{eff}}$(MARCS05s)) with the negative case of this equation corresponding to [Fe/H]$<-0.5$.  Signs are shown to indicate the direction of the change, but the error is the absolute value.  The error bars shown for the sun are if it is treated as all other stars, despite that a few of the errors are much less for this case. }  
\label{tab:stars}
\vspace{.5cm}
\begin{tabular}{cccrccrrcrrc}
\hline
\multicolumn{2}{c}{Star} & $\log g$ & \multicolumn{1}{c}{[Fe/H]} & Sources & $T_{\mathrm{eff}}$ & \multicolumn{1}{c}{$\Delta T_{\mathrm{eff}}$} &  \multicolumn{1}{c}{$\Delta T_{\mathrm{eff}}$} &$T_{\mathrm{eff}}$ & \multicolumn{1}{c}{$\Delta T_{\mathrm{eff}}$} &  \multicolumn{1}{c}{$\Delta T_{\mathrm{eff}}$} &  $T_{\mathrm{eff}}$\\
 HR & HD & \multicolumn{2}{c}{adopted}  & & H$\alpha$ & \multicolumn{1}{c}{H$\alpha$} & \multicolumn{1}{c}{H$\alpha$}  & H$\beta$  & \multicolumn{1}{c}{H$\beta$} & \multicolumn{1}{c}{H$\beta$} & adopted\\
 & & & & & & \multicolumn{1}{c}{conv} & \multicolumn{1}{c}{abund}  &  & \multicolumn{1}{c}{conv} & \multicolumn{1}{c}{abund} &  \\
 & &(cgs) & & & (K) & \multicolumn{1}{c}{(K)} &  \multicolumn{1}{c}{(K)} &  (K) &  \multicolumn{1}{c}{(K)} &  \multicolumn{1}{c}{(K)}  & (K) \\
\hline   
& & & & & & & & & & & \\ 
  Sun &         &   4.44 & $  0.00$ &           & 5733$\pm$ 81 & $ -11$ & $  39$ &      5723$\pm$ 90 & $  19$ & $   9$ &      5730$\pm$ 70 \\
  321 &    6582 &   4.39 & $ -0.86$ & AFG, AFG  & 5449$\pm$100 & $   6$ & $ -58$ &      5288$\pm$135 & $  57$ & $ -24$ &      5390$\pm$ 90 \\
  753 &   16160 &   4.50 & $ -0.08$ & AAM, AAM  & 4843$\pm$124 & $   3$ & $  52$ &                   &        &        &      4840$\pm$130 \\
      &   19445 &   4.50 & $ -1.88$ & AP, GCC   & 5783$\pm$172 & $ -29$ & $ -18$ &      5849$\pm$121 & $  33$ & $  -1$ &      5830$\pm$100 \\
  996 &   20630 &   4.50 & $  0.12$ & BLG, BLG  & 5733$\pm$103 & $  -9$ & $  38$ &      5682$\pm$102 & $  19$ & $   7$ &      5710$\pm$ 80 \\
 1101 &   22484 &   4.02 & $ -0.11$ & AP, GCC   & 6033$\pm$ 79 & $ -16$ & $  24$ &      6008$\pm$107 & $  40$ & $  -3$ &      6020$\pm$ 70 \\
      &   22879 &   4.38 & $ -0.76$ & AP, GCC   & 5867$\pm$119 & $  -5$ & $ -45$ &      5765$\pm$126 & $  45$ & $ -13$ &      5820$\pm$ 90 \\
 1325 &   26965 &   4.65 & $ -0.21$ & BG, BG    & 5187$\pm$111 & $   1$ & $  57$ &                   &        &        &      5190$\pm$120 \\
 2852 &   58946 &   4.17 & $ -0.31$ & SH, SH    & 6959$\pm$ 73 & $  -5$ & $  -6$ &      7006$\pm$184 & $ 127$ & $ -12$ &      6970$\pm$ 70 \\
 2943 &   61421 &   3.96 & $ -0.06$ & APP, AP   & 6538$\pm$ 82 & $ -22$ & $   4$ &      6474$\pm$143 & $  82$ & $ -13$ &      6520$\pm$ 80 \\
      &   64090 &   4.66 & $ -1.60$ & AP, GCC   & 5342$\pm$143 & $   4$ & $ -54$ &      5175$\pm$166 & $  85$ & $ -24$ &      5270$\pm$110 \\
 3775 &   82328 &   3.61 & $ -0.01$ & AP, AAM   & 6302$\pm$101 & $ -23$ & $   9$ &      6205$\pm$117 & $  53$ & $  -6$ &      6260$\pm$ 80 \\
 4150 &   91752 &   3.84 & $ -0.21$ & GCC, GCC  & 6567$\pm$116 & $ -24$ & $   1$ &      6404$\pm$135 & $  69$ & $  -9$ &      6500$\pm$ 90 \\
 4540 &  102870 &   3.98 & $  0.10$ & AP, GCC   & 6140$\pm$100 & $ -15$ & $  19$ &      6021$\pm$ 98 & $  26$ & $  -1$ &      6080$\pm$ 70 \\
 4550 &  103095 &   4.68 & $ -1.22$ & AP, GCC   & 5113$\pm$108 & $  17$ & $ -59$ &                   &        &        &      5110$\pm$110 \\
 4623 &  105452 &   4.00 & $ -0.60$ & AAM, AAM  & 7029$\pm$ 97 & $ -11$ & $   6$ &      6996$\pm$156 & $  96$ & $   8$ &      7020$\pm$ 90 \\
      &  108177 &   4.47 & $ -1.55$ & AP, GCC   & 6071$\pm$171 & $  39$ & $ -11$ &      6087$\pm$148 & $  75$ & $   3$ &      6080$\pm$120 \\
 4983 &  114710 &   4.32 & $  0.00$ & AP, GCC   & 6025$\pm$ 80 & $ -15$ & $  32$ &      5971$\pm$101 & $  37$ & $   2$ &      6000$\pm$ 70 \\
      &  114762 &   4.22 & $ -0.67$ & AP, GCC   & 5902$\pm$125 & $ -14$ & $ -39$ &      5806$\pm$131 & $  52$ & $  -4$ &      5860$\pm$100 \\
 5072 &  117176 &   3.80 & $ -0.11$ & BLG, BLG  & 5547$\pm$ 74 & $  -1$ & $  34$ &      5512$\pm$ 76 & $   0$ & $   8$ &      5530$\pm$ 60 \\
 5338 &  124850 &   3.52 & $ -0.11$ & AP, GCC   & 6202$\pm$ 81 & $ -22$ & $   7$ &      6115$\pm$100 & $  32$ & $  -4$ &      6170$\pm$ 70 \\
 5447 &  128167 &   4.21 & $ -0.41$ & AP, GCC   & 6769$\pm$111 & $ -13$ & $  -2$ &      6684$\pm$158 & $  93$ & $ -11$ &      6740$\pm$100 \\
      &  140283 &   3.81 & $ -2.40$ & AP, GCC   & 5589$\pm$177 & $  52$ & $  -7$ &      5677$\pm$155 & $  84$ & $   0$ &      5640$\pm$120 \\
 5868 &  141004 &   4.16 & $ -0.04$ & AP, GCC   & 5953$\pm$ 94 & $ -14$ & $  29$ &      5860$\pm$ 93 & $  20$ & $   3$ &      5910$\pm$ 70 \\
 5914 &  142373 &   3.99 & $ -0.46$ & AP, GCC   & 5817$\pm$ 90 & $ -17$ & $  37$ &      5726$\pm$105 & $  39$ & $   8$ &      5780$\pm$ 70 \\
 5933 &  142860 &   4.12 & $ -0.18$ & AP, GCC   & 6372$\pm$109 & $ -22$ & $  12$ &      6222$\pm$129 & $  65$ & $  -6$ &      6310$\pm$ 90 \\
 6775 &  165908 &   4.17 & $ -0.54$ & AP, GCC   & 6079$\pm$123 & $ -20$ & $ -25$ &      5901$\pm$113 & $  37$ & $  -3$ &      5980$\pm$ 90 \\
 6850 &  168151 &   3.93 & $ -0.32$ & AP, GCC   & 6413$\pm$112 & $ -25$ & $   7$ &      6343$\pm$140 & $  76$ & $  -8$ &      6390$\pm$ 90 \\
 7469 &  185395 &   4.40 & $ -0.02$ & BLG, BLG  & 6677$\pm$ 95 & $  -8$ & $   4$ &      6649$\pm$158 & $  99$ & $ -14$ &      6670$\pm$ 90 \\
 7534 &  187013 &   4.04 & $ -0.16$ & AP, GCC   & 6378$\pm$116 & $ -22$ & $  10$ &      6314$\pm$141 & $  76$ & $  -9$ &      6350$\pm$ 90 \\
 8832 &  219134 &   4.50 & $  0.00$ & AAM, AAM  & 4994$\pm$ 84 & $   3$ & $  50$ &                   &        &        &      4990$\pm$ 90 \\
\hline
\end{tabular}
\end{center}
Sources for $\log g$ and [Fe/H] parameters: AAM = Alonso~et~al.~(\cite{alonso}), AFG = Axer~et~al.~(\cite{axer1}), AP = Allende Prieto~et~al.~(\cite{allende_grav}), APP = Allende Prieto~et~al.~(\cite{allende_procyon}), BG = Bell \& Gustafsson~(\cite{bg1}), BLG = Blackwell \& Lynas-Gray~(\cite{blg94}), GCC = Gratton~et~al.~(\cite{gratton1}), SH = Saxner \& Hammarb\"ack~(\cite{saxner}) 
\end{minipage}
\end{table*}

\section{Comparisons and discussion}

\begin{figure*}
\begin{tabular}{cc}
\figpswh{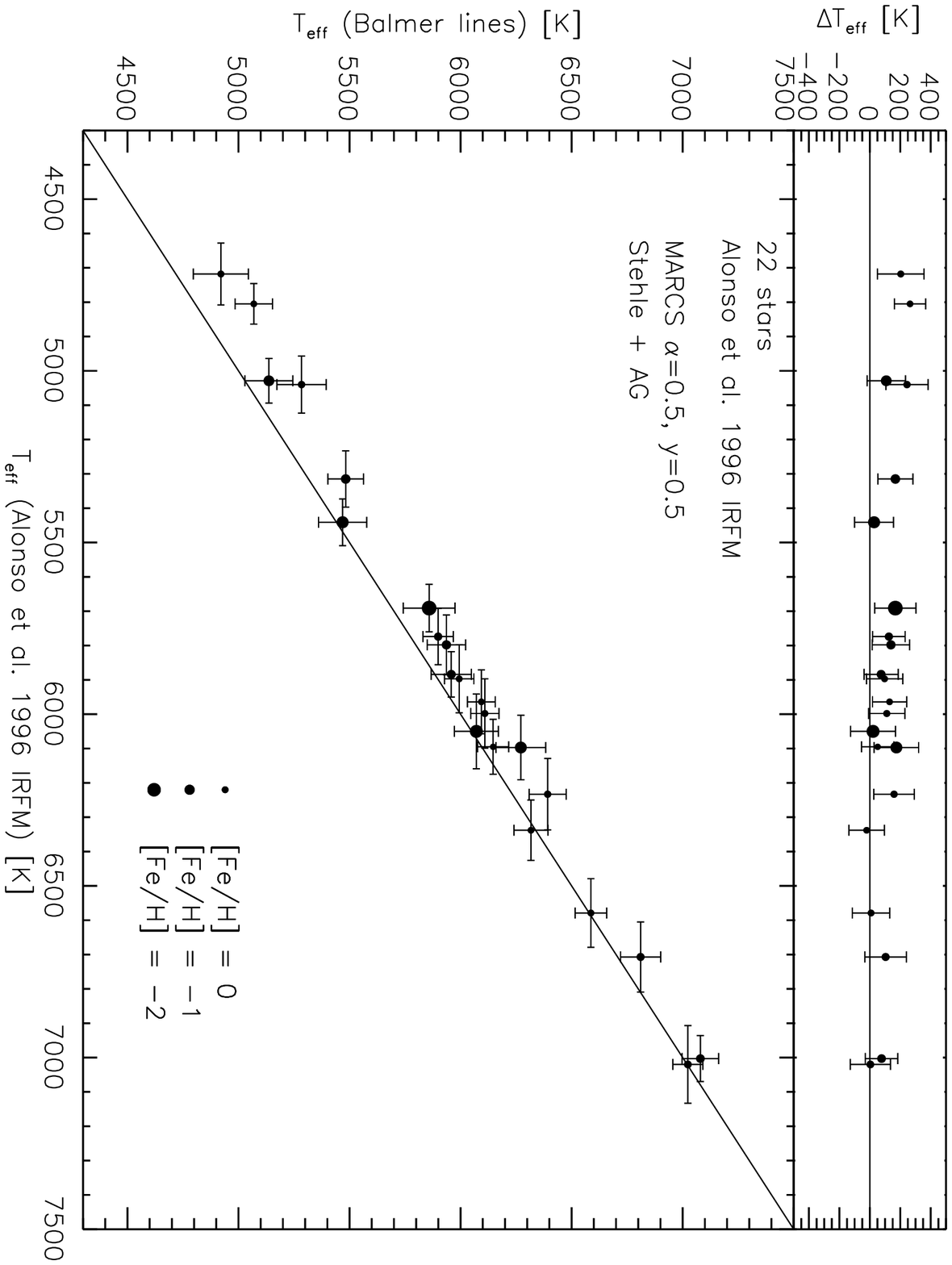} &
\figpswh{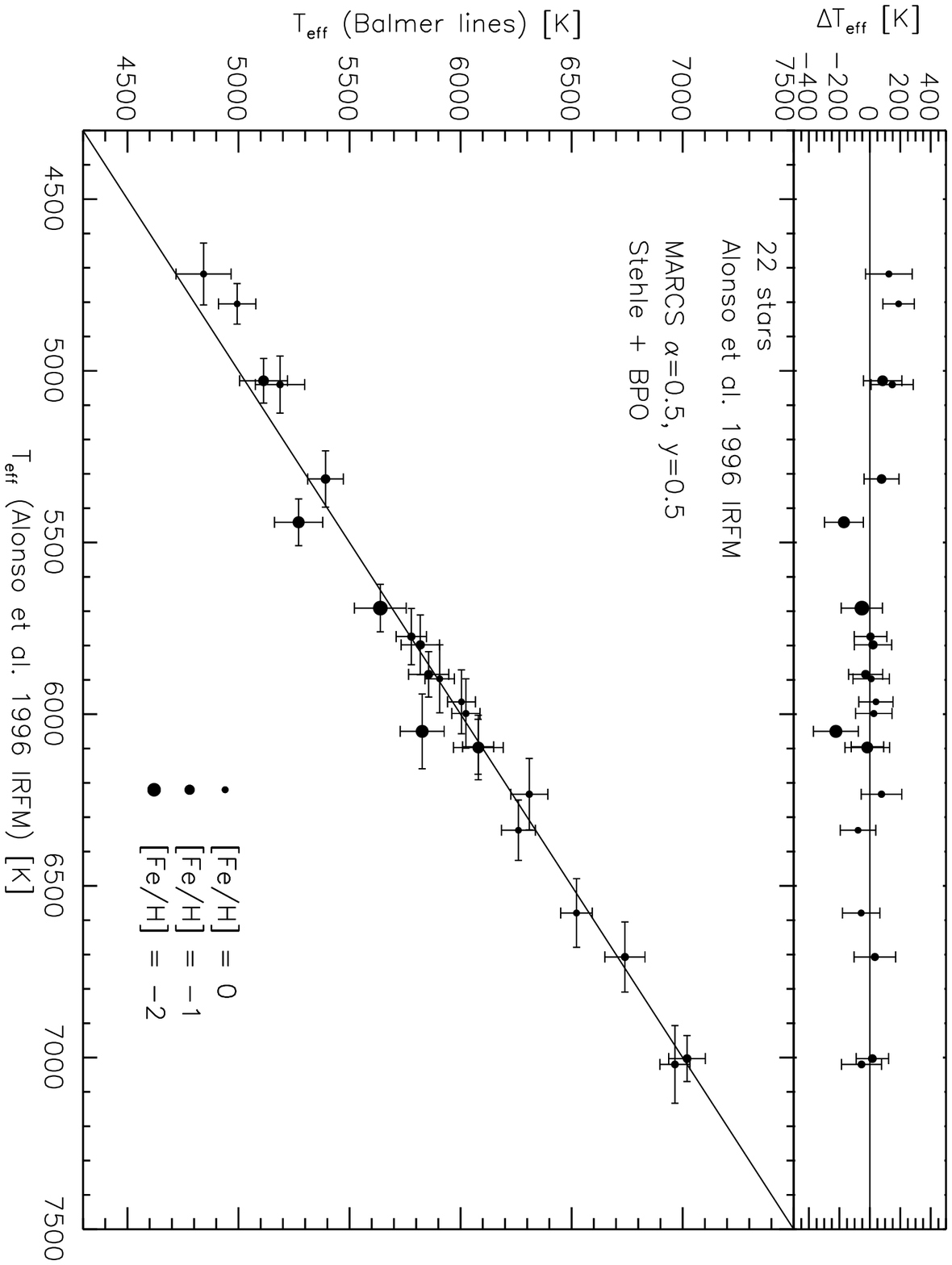}
\end{tabular}
\caption{Derived temperatures for STEHLE+AG (left) and STEHLE+BPO (right) recipes compared with the IRFM results of Alonso et~al.(\cite{alonso}). }
\label{fig:alonso}
\end{figure*}

Fig.~\ref{fig:alonso} compares our adopted effective temperatures against the results from the IRFM of Alonso~et~al.~(\cite{alonso}) for both STEHLE+AG and STEHLE+BPO recipes. We choose comparison with Alonso~et~al.~(\cite{alonso}) as their results include both solar metallicity and metal-poor stars, and has the largest overlap with our sample.  Using the STEHLE+AG recipe we find that the temperatures are on average hotter than those found from the IRFM by 110$\pm$77~K.  When the STEHLE+BPO recipe is used we see generally agreement within error, the results typically hotter by 11$\pm$95~K. Two outliers are seen, namely HD~19445 and HD~64090, which we find to be significantly cooler than the IRFM values from Alonso~et~al.~(\cite{alonso}). If we compare only those 4 stars in our sample with metallicity below $-$1.5, which includes these two stars, with Alonso~et~al.~(\cite{alonso}) a difference of $-$117$\pm$97~K is found.  The only other IRFM calibration we are aware of in the literature for metal-poor stars is by Magain~(\cite{magain}), who finds temperatures typically 112$\pm$56~K cooler than Alonso~et~al.~(\cite{alonso}) for 11 metal-poor stars.  Alonso~et~al.~(\cite{alonso}) attribute this zero point difference to the atmospheric models used.  If we again consider only the 4 stars in our sample with metallicity below $-$1.5 and compare with the results of Magain~(\cite{magain}) the difference is $-$34$\pm$85~K, our temperatures from Balmer lines being slightly cooler. These 4 stars are the only common stars with the Magain~(\cite{magain}) sample. Thus, our results are in better agreement with Magain~(\cite{magain}) for the metal-poor stars, though 4 stars is insufficient for any clear conclusion.

There is a suggestion from Fig.~\ref{fig:alonso} that our temperatures are systematically hotter than Alonso~et~al.~(\cite{alonso}) in the coolest stars in the sample.  This may well be due to the increased blending causing an overestimate of the Balmer line temperatures, though we note the IRFM may also be less reliable in this regime.

Similar comparisons were made for common stars with the results for population I stars of Blackwell \&\ Lynas-Gray~(\cite{blg94,blg98}) and Saxner \&\ Hammerb\"ack~(\cite{saxner}) both using the IRFM and the results showed generally agreement within error.   Agreement with $T_{\mathrm{eff}}$ values for 16 common stars from Edvardsson~et~al.~(\cite{bdp}) using $b-y$ colours was also good, with a mean difference of $-51\pm67$~K.  Two exceptions were Procyon and HR 2852 both early F stars, which we found to be significantly cooler than the Edvardsson~et~al.~(\cite{bdp}) temperatures.
 
\begin{figure*}
\begin{tabular}{cc}
\figpswh{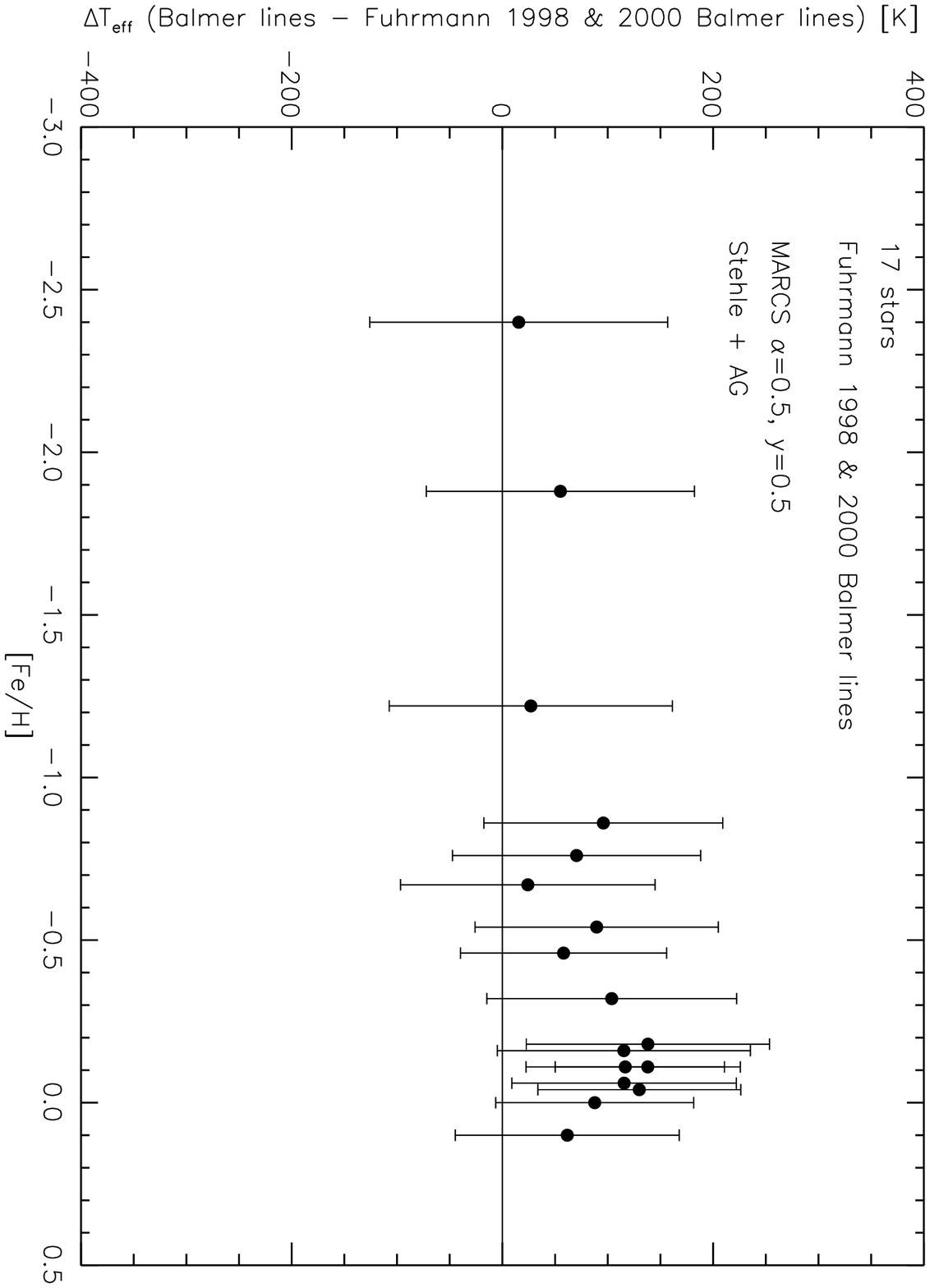} &
\figpswh{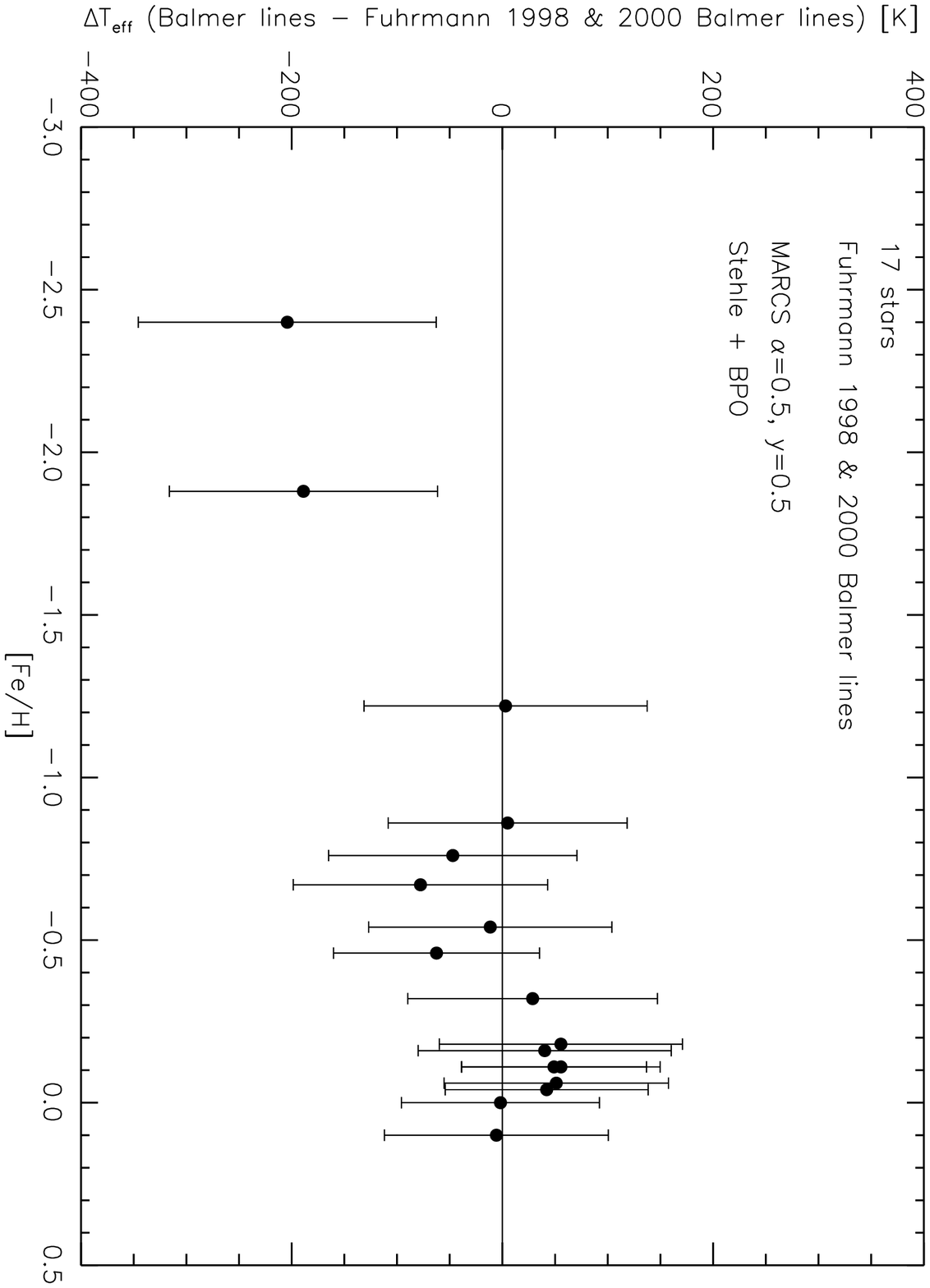}
\end{tabular}
\caption{Derived temperatures for STEHLE+AG (left) and STEHLE+BPO (right) recipes compared with the Balmer line results of Fuhrmann against metallicity.}
\label{fig:fuhrmanncomp}
\end{figure*}

Fuhrmann~et~al.~(\cite{fuhrmann94}) determined temperatures for a large sample of stars using Balmer lines, these determinations having been superseded by more recent work with better observations and improved models (Fuhrmann~\cite{fuhrmann98,fuhrmann00}).  We compare our temperatures with Fuhrmann's work (\cite{fuhrmann98,fuhrmann00}) in Fig.~\ref{fig:fuhrmanncomp} (right panel).  As Fuhrmann employs the Ali \& Griem~(\cite{ali_griem:errata}) theory we might have expected to see differences of the order predicted in paper II (Fig. 7 of that paper). In fact, the agreement around solar metallicity and even slightly below is good.  Comparison of our results using the STEHLE+AG recipe with those of Fuhrmann are shown in Fig.~\ref{fig:fuhrmanncomp} (left panel).  Reasonable agreement between the two sets of results might have been expected, but the results show a systematic difference of 91$\pm$40~K with no significant trend with stellar parameters.  The reason for this difference has been traced to a number of factors, namely systematic differences in Stark broadening, model atmospheres, and a slightly different application of the Ali \&\ Griem~(\cite{ali_griem:errata}) theory (see Sect.~\ref{sect:models_and_spectra}).  These differences are all quite small but by chance all act in the same direction, with relative importance varying with stellar parameters and line.  The differences are also blurred by random errors such as the use of different observations, fitting methods and adopted gravity and metallicity.  For example using MARCS05 and STEHLE+AG we fit the solar H$\alpha$ spectra best with $T_\mathrm{eff}\approx5830$~K, while Fuhrmann~et~al.~(\cite{fuhrmann97}) fit the line with  $T_\mathrm{eff}\approx5750$~K.  As shown in Sect.~\ref{sect:compmodels}, $\approx20$~K of this originates in the atmospheric models.  Comparisons of the Stark broadening tables employed by Fuhrmann~et~al. were found to give very slightly stronger profiles than VCS, and thus an even greater difference when compared to STEHLE than VCS.  This accounts for a further $\approx10$~K (Table~\ref{tab:errors}).  If one adds the resonance broadening of the 3p state to that of the 2p state, the broadening is increased by about 15$\%$ compared to considering only the 2p state as we do in our STEHLE+AG recipe.  We see from Table~\ref{tab:errors} that this produces a significant difference of $\approx40$~K.  As stated above all these differences act in the same sense in that Fuhrmann~et~al.'s profiles are stronger, and thus explain the difference of $\approx80$~K in derived temperatures.  Similarly, for H$\beta$ we obtain $T_\mathrm{eff}\approx5820$~K, and Fuhrmann~et~al.~$T_\mathrm{eff}\approx5750$~K.  For this line the difference in models and implementation of resonance broadening is less important, $\approx10$~K each, while the Stark broadening difference is increased $\approx20$~K, explaining $\approx40$~K of the difference.  The remainder is certainly within fitting and observational error.  We note that these differences of order 80~K between the two determinations are consistent with, and thus support, our error estimates.     

\section{Conclusions}

Effective temperatures have been derived from Balmer line profiles using a quantitative fitting method with a detailed error analysis including investigation of the susceptibility to various errors with temperature and metallicity for our models.  Our temperatures find good agreement with the IRFM near solar metallicity but show differences at low metallicity where the two available IRFM determinations themselves are in disagreement.  Our results for metal-poor stars seem in better agreement with Magain~(\cite{magain}), though we caution this is only for 4 stars and the better agreement may be fortuitous.  This should be investigated further, for which more, and better observations of Balmer lines in population II stars are needed.   The origin of the differences between these two IRFM determinations needs also to be understood.

Through estimates of the uncertainties, we found that the relative weight that should be given to H$\alpha$ and H$\beta$ \emph{in determining effective temperatures} varies quite substantially with stellar parameters.  This is predominantly a matter of balance between the reduced sensitivity of H$\alpha$ to temperature at low metallicities making broadening, $\log g$ and observational uncertainties very important, and the high sensitivity of H$\beta$ to convection.  Thus to improve the accuracy of Balmer line temperatures for metal-poor stars these four areas must be addressed.  Observational uncertainties will continue to improve as large telescopes provide the possibility of combined high resolution and SNR for these faint objects.  There no doubt exists a limiting accuracy for continuum determination procedures for echelle spectra such as that used here, but we believe this has not yet been reached.  Fuhrmann~et~al.~(\cite{fuhrmann94}) suggest an accuracy of about 0.3$\%$ is achievable with modern spectroscopy.  Planned satellite astrometry missions should address the gravities.  Both the remaining uncertainties, convection and self-broadening, lie in the theoretical realm at least at present.  Considerable progress has been made in 3D hydrodynamical simulations of convection and inhomogeneities in cool stars, and this will be the way forward.  However, for the immediate future where direct application of such models is impractical, calibrations of MLT parameters \emph{for Balmer lines} across the HR diagram similar to Ludwig~et~al.~(\cite{ludwig}) or tabulated $T_\mathrm{eff}$ corrections (against a given MLT parameter set) would be important.  Improved calculations of the self-broadening of Balmer lines should be undertaken without resort to the impact approximation which will require improved short range potentials from those used in paper II.

In paper I and II large differences between $T_\mathrm{eff}$ values derived with Ali \& Griem~(\cite{ali_griem:errata}) theory and new calculations of the self-broadening, particularly in metal-poor stars, were predicted.  Comparison of our results with those of Fuhrmann~(\cite{fuhrmann98,fuhrmann00}) where Ali \& Griem theory has been employed, find in fact good agreement except for low metallicity stars of around solar temperature where the differences are smaller than expected, but the temperatures are still significantly cooler. The reason for this unexpected agreement is a number of systematic differences in models and employed broadening recipes, which though typically individually small, together somewhat compensate the difference made by the new self-broadening. This emphasises that in order to achieve high absolute precision in $T_\mathrm{eff}$ determinations from Balmer lines such small differences must in fact be carefully considered.  Errors in the relative temperature scale, particularly between stars of low and solar metallicity important in tracing chemical evolution in the galaxy, are unlikely to be much smaller than the errors in the absolute scale since those errors which might cancel in differential comparison, namely broadening theory and model errors (including convection), typically vary with metallicity resulting in little cancellation.

\begin{acknowledgements}

We thank the stellar atmospheres group at Uppsala for valuable discussions and encouragement. David Lambert is thanked for acquiring the spectrum of Gmb 1830 and for hospitality during a visit by PB to University of Texas, Austin.  Andreas Korn is thanked for valuable discussions and co-operation, and providing MUNICH models. We thank the referee, Prof. T. Gehren, for valuable comments.  PB acknowledges the support of the Swedish Natural Science Research Council and Arvid Sch\"onberg stipend. CA acknowledges support from the US National Science Foundation (grant AST-0086321). NSO/Kitt Peak FTS data used here were produced by NSO/NOAO.

\end{acknowledgements}

\end{document}